\documentclass{aa}

\usepackage{graphicx}
\usepackage{amsmath}	
\usepackage{amssymb}	
\usepackage{txfonts}
\usepackage{natbib}
\usepackage{longtable}
\usepackage[normalem]{ulem}
\usepackage{tablefootnote}

\newcommand{\teff}{${T}_{\mathrm{eff}}$}
\newcommand{\logg}{$\log{g}$}
\newcommand{\msun}{$M_{\odot}$}

\begin{document}

   \title{\emph{TESS} first look at evolved compact pulsators}

   \subtitle{Asteroseismology of the pulsating helium-atmosphere white dwarf TIC\,257459955}

   \author{Keaton~J.~Bell\inst{1,2}\fnmsep\thanks{NSF Astronomy and Astrophysics Postdoctoral Fellow \newline Current address: DIRAC Institute, Department of Astronomy, University of Washington,
  Seattle, WA 98195-1580, USA \newline \email{keatonb@uw.edu}},
        Alejandro H.~C\'orsico\inst{3,4},
        Agn\`es Bischoff-Kim\inst{5},
        Leandro G.~Althaus\inst{3,4},
        P.~A.~Bradley\inst{6},
        Leila~M.~Calcaferro\inst{3,4},
        M.~H.~Montgomery\inst{7},
        Murat Uzundag\inst{8},
        Andrzej S. Baran\inst{9},
        Zs.~Bogn\'ar\inst{10,11},
        S.~Charpinet\inst{12},
        H.~Ghasemi\inst{13},
        \and J.~J.~Hermes\inst{14}}
   
   \institute{Max-Planck-Institut f{\"u}r Sonnensystemforschung (MPS), Justus-von-Liebig-Weg 3, 
		37077 G{\"o}ttingen, Germany
		\and
		Department of Physics and Astronomy, Stellar Astrophysics Centre, Aarhus University, Ny Munkegade 120, 8000 Aarhus C, Denmark
		\and
		Grupo de Evoluci\'on Estelar y Pulsaciones. 
           Facultad de Ciencias Astron\'omicas y Geof\'{\i}sicas, 
           Universidad Nacional de La Plata, 
           Paseo del Bosque s/n, 1900 
           La Plata, 
           Argentina
        \and
        IALP - CONICET
        \and
        Penn State Worthington Scranton, Dunmore, PA 18512, USA
        \and
        XCP-6, MS F-699 Los Alamos National Laboratory, Los Alamos, NM 87545, USA
        \and
        Department of Astronomy, University of Texas at Austin, Austin, TX\,-\,78712, USA
        \and
           Instituto de F\'isica y Astronom\'ia, Universidad de Valparaiso, Gran Breta\~na 1111, Playa Ancha, Valpara\'iso 2360102, Chile
        \and
         Uniwersytet Pedagogiczny, Obserwatorium na Suhorze, ul. Podchor\c{a}\.zych 2, 30-084 Krak\'ow, Polska
        \and
        Konkoly Observatory, MTA Research Centre for Astronomy and Earth Sciences, Konkoly Thege Mikl\'os \'ut 15-17, H--1121, Budapest, Hungary
        \and
        MTA CSFK Lend\"ulet Near-Field Cosmology Research Group
        \and
        Institut de Recherche en Astrophysique et Plan{\'e}tologie,
        CNRS, Universit{\'e} de Toulouse, CNES, 14 avenue Edouard Belin, F-31400 Toulouse, France
        \and
        Department of Physics, Institute for Advanced Studies in Basic Sciences (IASBS), Zanjan 45137-66731, Iran
        \and
        Department of Astronomy, Boston University, 725 Commonwealth Ave., Boston, MA 02215, USA
        }

   \date{}

    \titlerunning{Asteroseismology of the DBV TIC\,257459955 with \emph{TESS}}
	\authorrunning{Keaton J.~Bell et al.}

  \abstract
   {Pulsation frequencies reveal the interior structures of white dwarf stars, shedding light on the properties of these compact objects that represent the final evolutionary stage of most stars. Two-minute cadence photometry from the Transiting Exoplanet Survey Satellite (\emph{TESS}) will record pulsation signatures from bright white dwarfs over the entire sky.}
   {As part of a series of first-light papers from \emph{TESS} Asteroseismic Science Consortium Working Group 8, we aim to demonstrate the sensitivity of \emph{TESS} data to measuring pulsations of helium-atmosphere white dwarfs in the DBV instability strip, and what asteroseismic analysis of these measurements can constrain about their stellar structures.  We present a case study of the pulsating DBV WD\,0158$-$160 that was observed as TIC\,257459955 with the 2-minute cadence for 20.3 days in \emph{TESS} Sector 3.}
   {We measure the frequencies of variability of TIC\,257459955 with an iterative periodogram and prewhitening procedure. The measured frequencies are compared to calculations from two sets of white dwarf models to constrain the stellar parameters: the fully evolutionary models from {\tt LPCODE}, and the structural models from {\tt WDEC}.}
   {We detect and measure the frequencies of nine pulsation modes and eleven combination frequencies of WD\,0158$-$160 to $\sim0.01\,\mu$Hz precision. Most, if not all, of the observed pulsations belong to an incomplete sequence of dipole ($\ell=1$) modes with a mean period spacing of $38.1\pm1.0$\,s. The global best-fit seismic models from both {\tt LPCODE} and {\tt WDEC} have effective temperatures that are $\gtrsim3000$\,K hotter than archival spectroscopic values of $24{,}100$--$25{,}500$\,K; however, cooler secondary solutions are found that are consistent with both the spectroscopic effective temperature and distance constraints from \emph{Gaia} astrometry.}
   {Our results demonstrate the value of the \emph{TESS} data for DBV white dwarf asteroseismology. 
   The extent of the short-cadence photometry enables reliably accurate and extremely precise pulsation frequency measurements. 
   Similar subsets of both the {\tt LPCODE} and {\tt WDEC} models show good agreement with these measurements, supporting that the asteroseismic interpretation of DBV observations from \emph{TESS} is not dominated by the set of models used; however, given the sensitivity of the observed set of pulsation modes to the stellar structure, external constraints from spectroscopy and/or astrometry are needed to identify the best seismic solutions.
   }


   \maketitle

\section{Introduction}

\begin{figure*}
	\centering
	\includegraphics[width=1.9\columnwidth]{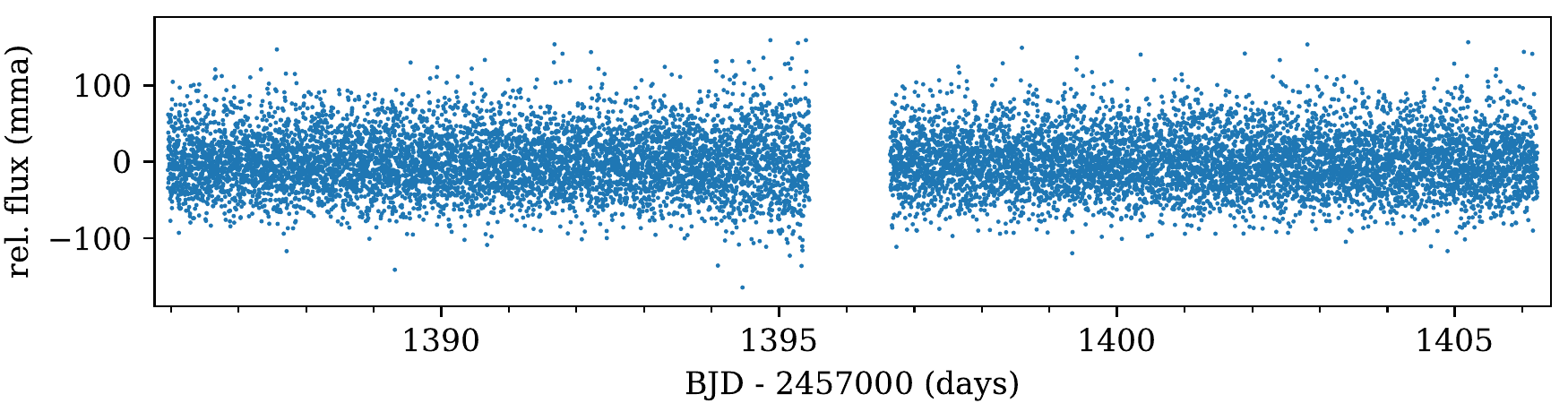}
	\caption{Final reduced \emph{TESS} Sector 3 light curve of TIC\,257459955. A gap occurs at spacecraft perigee.} 
	\label{fig:lc}
\end{figure*}

The Transiting Exoplanet Survey Satellite (\emph{TESS}) is a NASA mission with the primary goal of detecting exoplanets that transit the brightest and nearest stars \citep{Ricker2014}. More generally, the extensive time series photometry that \emph{TESS} acquires is valuable for studying a wide variety of processes that cause stars to appear photometrically variable.  One particularly powerful use for these data is to constrain the global properties and interior structures of pulsating stars with the methods of asteroseismology. Pulsating stars oscillate globally in standing waves that propagate through and are affected by the stellar interiors. Fourier analysis of the light curves of pulsating stars reveals their eigenfrequencies that can be compared to calculations from stellar models, providing the most sensitive technique for probing stellar interior structures.

The \emph{TESS} Asteroseismic Science Consortium (TASC) is a collaboration of the scientific community that shares an interest in utilizing \emph{TESS} data for asteroseimology research. It is organized into a number of working groups that address different classes of stars. TASC Working Group 8 (WG8) focuses on \emph{TESS} observations of evolved compact stars that exhibit photometric variability, including hot subdwarfs, white dwarf stars, and pre-white dwarfs. To this goal, WG8 has proposed for all known and likely compact stars with \emph{TESS} magnitudes $\lesssim16$ to be observed at the short, 2-minute cadence.

Within TASC WG8, the subgroup WG8.2 coordinates the studies of pulsating white dwarfs observed by TESS. Depending on their atmospheric compositions, white dwarfs may pulsate as they cool through three distinct instability strips: DOVs (GW Vir stars or pulsating PG\,1159 stars) are the hottest and include some central stars of planetary nebulae; DBVs (V777\,Her stars) have helium atmospheres that are partially ionized in the effective temperature range $32{,}000 \gtrsim T_\mathrm{eff} \gtrsim 22{,}000$\,K, driving pulsations; and DAVs (ZZ Ceti stars) pulsate when their pure-hydrogen atmospheres are partially ionized from $12{,}500 \gtrsim T_\mathrm{eff} \gtrsim 10{,}800$\,K (at the canonical mass of $\approx$\,0.6\,\msun). Pulsations of these objects probe the physics of matter under the extreme pressures of white dwarf interiors.  Since white dwarfs are the final products of $\approx$97\% of Galactic stellar evolution, asteroseismic determination of their compositions and structures probes the physical processes that operate during previous evolutionary phases. See \citet{Winget2008}, \citet{Fontaine2008}, and \citet{Althaus2010} for reviews of the field of white dwarf asteroseismology, and \citet{Corsico2019} for coverage of the most recent decade of discovery in the era of extensive space-based photometry from \emph{Kepler} and K2.

As part of the initial activities of the TASC WG8.2, we present analyses of examples of each type of pulsating white dwarf observed at 2-minute cadence in the first \emph{TESS} Sectors in a series of first light papers. These follow the TASC WG8.3 first-light analysis of a pulsating hot subdwarf in \emph{TESS} data from Charpinet et al.\ (2019, submitted). In this paper, we study the DBV pulsator WD\,0158$-$160 (also EC\,01585$-$1600, G\,272-B2A), which was observed by \emph{TESS} as target TIC\,257459955 in Sector 3. \citet{Voss2007} confirm the classification of WD\,0158$-$160 as a DB (helium-atmosphere) white dwarf from a ESO Supernova type Ia Progenitor surveY (SPY) spectrum and measure atmospheric parameters of \teff\ $=25{,}518$\,K and \logg\ $=7.875$. 
The more recent spectroscopic study of \citet{Rolland2018} finds a cooler best-fit model to their observations, obtaining \teff\ $=24{,}130\pm1369$\,K and \logg\ $=7.94\pm0.03$.
Astrometric parallax from \emph{Gaia} DR2 \citep{Gaia2016,Gaia2018} place WD\,0158$-$160 at a distance of $68.14\pm0.28$\,pc \citep{BailerJones2018}.  This is one of the brightest DBVs known \citep[$V=14.55\pm0.08$\,mag;][]{Zacharias2012} and was discovered to be a variable by \citet{Kilkenny2016}. They obtained high-speed photometry on the Sutherland 1-meter telescope of the South African Astronomical Observatory over five nights, measuring ten frequencies of significant variability between 1285--5747\,$\mu$Hz.  We aim to measure more precise pulsation frequencies from the \emph{TESS} data and to compare these with stellar models to asteroseismically constrain the properties of this DB white dwarf.

\section{\emph{TESS} data}

TIC\,257459955 was observed at the short, 2-minute cadence by \emph{TESS} in Sector 3, which collected 20.3 days of useful data with a 1.12-day gap at spacecraft perigee.\footnote{See \emph{TESS} Data Release Notes: \\ \url{http://archive.stsci.edu/tess/tess_drn.html}}  Light curves from this particular Sector are shorter than the nominal 27-day duration, and the periodogram achieves a correspondingly lower frequency resolution and signal-to-noise than expected for most \emph{TESS} observations.  Thus, TESS's value for asteroseismology of white dwarfs observed in Sectors with longer coverage is typically greater than demonstrated in this paper.

We use the 2-minute short-cadence \emph{TESS} light curve of TIC\,257459955 that has had common instrumental trends removed by the Pre-Search Data Conditioning Pipeline \citep[PDC;][]{Stumpe2012} that we downloaded from MAST.\footnote{\url{https://archive.stsci.edu/}}  We discard two observations that have quality flags set by the pipeline.  
We do not identify any additional outlying measurements that need to be removed. The final light curve contains 13{,}450 measurements that span 20.27 days.

\begin{figure*}
	\centering
	\includegraphics[width=1.9\columnwidth]{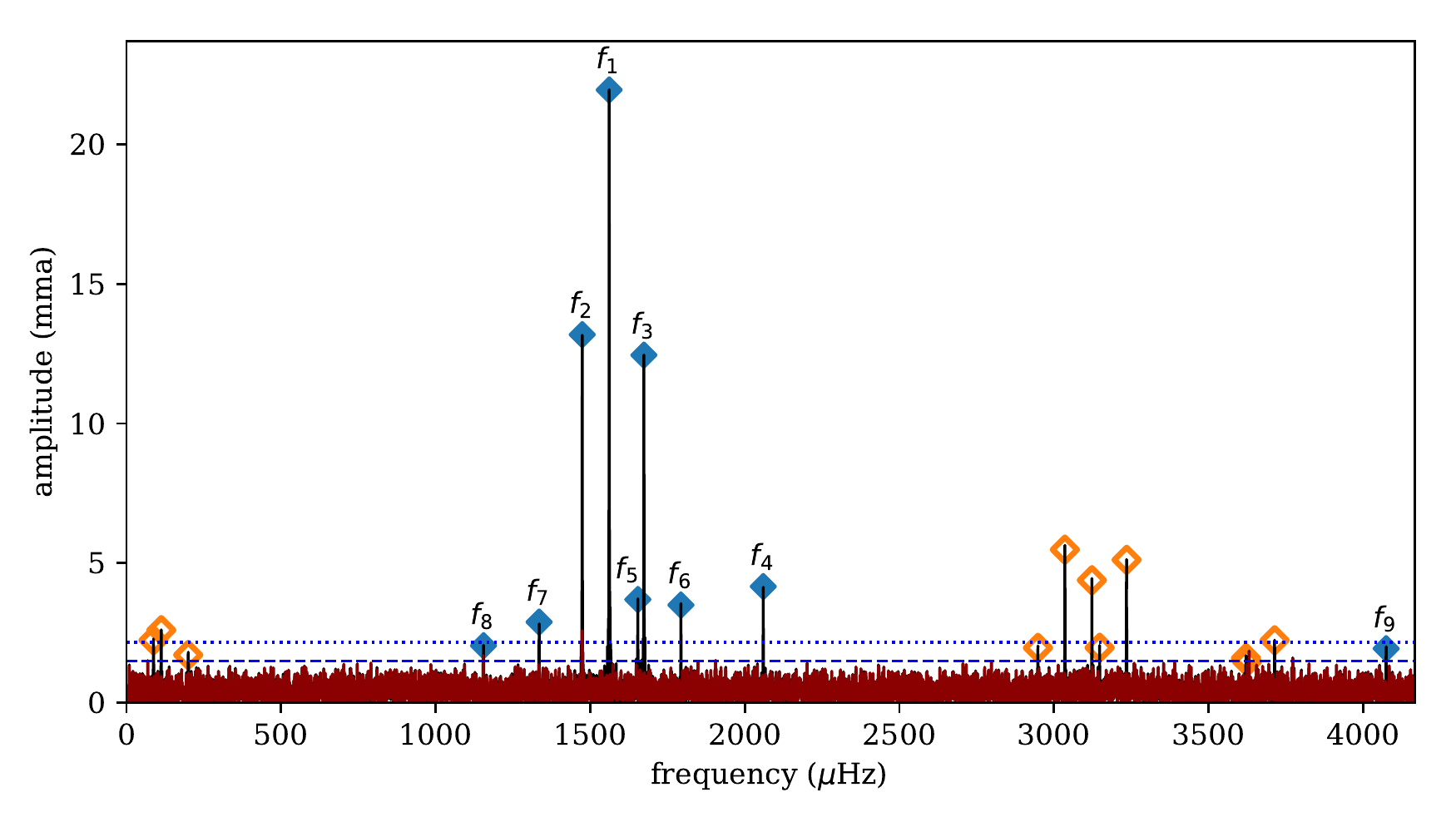}
	\caption{Lomb-Scargle periodogram of the original TIC\,257459955 light curve (black) and of the residuals after subtracting off the best fit frequency solution (red).  The dotted horizontal line shows the final 0.1\% false-alarm-probability (FAP) significance threshold for the residual spectrum, and the dashed line is a 0.1\% FAP level for an individual frequency bin. Blue filled diamonds mark the best-fit frequencies and amplitudes of independent modes, and orange, unfilled diamonds mark combination frequencies. Independent modes are labeled with mode IDs from Table~\ref{tab:freqs}.} 
	\label{fig:ft}
\end{figure*}

To remove any additional low-frequency systematics from the light curve, we divide out the fit of a fourth-order Savitzky–Golay filter with a three-day window length computed with the Python package {\sc lightkurve} \citep{lightkurve}.  This preserves the signals from pulsations that typically have periods of $\lesssim20$\,min in white dwarfs. The final reduced light curve is displayed in Figure~\ref{fig:lc}, where the relative flux unit of milli-modulation amplitude (mma) equals 0.1\% flux variation or one part-per-thousand. The root-mean-squared scatter of the flux measurements is 37.8\,mma (3.78\%).

\section{Frequency solution}\label{sec:frequencies}

Astroseismology relies on the precise determination of pulsation frequencies. We benefit from the length of the \emph{TESS} data providing a high frequency resolution of 0.57\,$\mu$Hz (inverse of the light curve duration) without complications from aliasing that would arise from large gaps in the time series. We use the fast Lomb-Scargle implementation in \verb+astropy+ \citep{astropy} to compute periodograms of the unweighted time series photometry. We oversample the natural frequency resolution by a factor of 10 so that the periodogram peaks more accurately represent the intrinsic frequencies and amplitudes of the underlying signals.  The full periodogram out to the Nyquist frequency of 4166.59\,$\mu$Hz is displayed in black in Figure~\ref{fig:ft}.

Including any noise peaks in our frequency solution would adversely affect our asteroseismic inferences, so we adopt a conservative significance criterion for signal amplitudes. We test the null hypothesis that the highest peak in the periodogram is caused by pure noise by treating the observed flux measurements in the light curve as a proxy for the noise distribution. We bootstrap a significance threshold by generating 10{,}000 pure-noise time series that sample from this distribution with replacement at the observation times of the original light curve. The 99.9th highest percentile corresponds to a false alarm probability (FAP) of 0.1\% that a peak with a higher amplitude anywhere in the oversampled periodogram is caused by noise alone. We have high confidence that peaks above this threshold correspond to significant signals. For our initial periodogram, we find that peaks with amplitudes above 2.71\,mma (4.7 times the mean noise level in the periodogram\footnote{This matches the significance threshold advocated for by \citet{Baran2015}, though they use a different method to arrive at this level, interpreting it as the threshold that yields the correct frequency determinations in 95\% of random realizations.}) have FAP $< 0.1$\%.

\begin{table*}
	\centering
	\caption{Pulsational frequency solution from the \emph{TESS} light curve of TIC\,257459955. The independent pulsation modes are listed first in order of decreasing amplitude, followed by all identified combination frequencies.}
	\label{tab:freqs}
	\begin{tabular}{cccc} 
		\hline
		mode & frequency & period & amplitude\\
		ID & ($\mu$Hz) & (s) & (mma) \\
		\hline
		$f_1$ & $1561.200 \pm 0.005$ &  $640.533 \pm 0.002$ & $ 22.0 \pm 0.4  $ \\
        $f_2$ & $1473.985 \pm 0.008$ &  $678.433 \pm 0.004$ & $ 13.2 \pm 0.4  $ \\
        $f_3$ & $1673.490 \pm 0.008$ &  $597.554 \pm 0.003$ & $ 12.5 \pm 0.4  $ \\
        $f_4$ & $2059.62 \pm 0.02$ &  $485.527 \pm 0.006$ & $ 4.2 \pm 0.4  $ \\
        $f_5$ & $1653.87 \pm 0.03$ &  $604.642 \pm 0.010$ & $ 3.7 \pm 0.4  $ \\
        $f_6$ & $1793.24 \pm 0.03$ &  $557.649 \pm 0.010$ & $ 3.5 \pm 0.4  $ \\
        $f_7$ & $1334.64 \pm 0.04$ &  $749.27 \pm 0.02$ & $ 2.9 \pm 0.4  $ \\
        $f_8$ & $1154.77 \pm 0.06$ &  $865.97 \pm 0.04$ & $ 2.0 \pm 0.4  $ \\
        $f_9$ & $4074.32 \pm 0.06$ &  $245.440 \pm 0.004$ & $ 1.9 \pm 0.4  $ \\
		\hline
		$f_1 - f_2$ & $87.215 \pm 0.009$ &  $11466.0 \pm 1.2$ & $ 2.23 \pm 0.4  $ \\
        $f_3 - f_1$ & $112.290 \pm 0.010$ &  $8905.5 \pm 0.8$ & $ 2.60 \pm 0.4  $ \\
        $f_3 - f_2$ & $199.504 \pm 0.011$ &  $5012.4 \pm 0.3$ & $ 1.71 \pm 0.4  $ \\
        $2 f_1$ & $3122.400 \pm 0.009$ &  $320.267 \pm 0.001$ & $ 4.39 \pm 0.4  $ \\
        $2 f_2$ & $2947.971 \pm 0.015$ &  $339.2164 \pm 0.0017$ & $ 1.97 \pm 0.4  $ \\
        $f_1 + f_2$ & $3035.185 \pm 0.009$ &  $329.4692 \pm 0.0009$ & $ 5.5 \pm 0.4  $ \\
        $f_1 + f_3$ & $3234.690 \pm 0.009$ &  $309.1487 \pm 0.0009$ & $ 5.1 \pm 0.4  $ \\
        $f_1 + f_4$ & $3620.82 \pm 0.02$ &  $276.1809 \pm 0.0018$ & $ 1.6 \pm 0.4  $ \\
        $f_2 + f_3$ & $3147.475 \pm 0.011$ &  $317.7150 \pm 0.0011$ & $ 2.0 \pm 0.4  $ \\
        $f_4 + f_5$ & $3713.49 \pm 0.03$ &  $269.289 \pm 0.002$ & $ 2.2 \pm 0.4  $ \\
        $f_1 + f_2 + f_3$ & $4708.675 \pm 0.012$ &  $212.3740 \pm 0.0005$ & $ 1.5 \pm 0.4  $ \\
		\hline
	\end{tabular}
\end{table*}

We adopt frequencies into our solution according to an iterative prewhitening procedure. We record the frequencies and amplitudes of every peak above our 0.1\% FAP significance threshold.  These provide initial values for a multi-sinusoid fit to the time series data,\footnote{Our interactive Python-based periodogram and sine-fitting code is available at \url{https://github.com/keatonb/Pyriod}.} which we compute with the non-linear least-squares minimization Python package \verb+lmfit+ \citep{lmfit}.  Frequencies that agree within the natural frequency resolution with a sum, integer multiple, or difference between higher-amplitude signals are identified as combination frequencies.  These arise from a nonlinear response of the flux to the stellar pulsations \citep{Brickhill1992}, and we enforce a strict arithmetic relationship between these combination frequencies and the parent pulsation frequencies when performing the fit. Including the combination frequencies slightly improves the measurement precision of the parent mode frequencies. Once all significant signals are included in our model and fitted to the time series, we subtract off the model and repeat the process on the residuals, starting by recalculating the periodogram and bootstrapping its 0.1\% FAP significance threshold. This is repeated until no further signals meet our acceptance criterion.

At this point, the periodogram still exhibits a few compelling peaks at locations where we specifically expect that signals might appear. For assessing significance in these cases, we bootstrap a different 0.1\% FAP threshold for the amplitude of a peak within a single frequency bin (as opposed to considering the highest peak anywhere in the entire spectrum).  
We adopt signals that correspond to combinations of accepted modes into our solution that exceed this lower threshold. We accept another independent pulsation mode at $1154.77\pm0.06\,\mu$Hz ($f_8$ in Table~\ref{tab:freqs}) that closely matches the asymptotic mean period spacing of $\ell=1$ modes that we identify in our preliminary asteroseismic mode identification in Section~\ref{sec:modeids}.
We also include the peak at $4074.32\pm0.06$\,$\mu$Hz ($f_9$) as an intrinsic pulsation mode since it agrees with the measurement of a frequency at $4074.0\pm0.1$\,$\mu$Hz from the ground-based discovery data of \citet{Kilkenny2016} to within 
the periodogram frequency resolution.
After prewhitening these signals, the final single-bin significance threshold is at 1.50\,mma,\footnote{The peak corresponding to the sub-Nyquist alias of the $f_1+f_2+f_3$ combination frequency exceeded this threshold even though the best-fit amplitude in the final solution is lower.} compared to the 0.1\% FAP level across the entire spectrum at 2.24\,mma. 

The final best-fit values for the frequencies (periods) and amplitudes of the individual sinusoids in our model are given in Table~\ref{tab:freqs}.  
The quoted errors are estimated by \verb+lmfit+ from the covariance matrix, and they agree with expectations from analytical formulae for the independent modes \citep{Montgomery1999}. In Figure~\ref{fig:ft}, the dotted line indicates the final full-spectrum 0.1\% FAP significance threshold and the dashed line marks the lower per-bin threshold.
These values are indicated by diamond markers (independent modes in filled blue and combination frequencies in unfilled orange).  
The periodogram of the final residuals is displayed in red. 
The measured amplitudes will generally be lower than the intrinsic disc-integrated amplitudes due to smoothing from the 2-minute exposures.  The intrinsic frequency of the combination $f_1+f_2+f_3$ is above the observational Nyquist frequency, so we mark the corresponding alias peak near 3624.5\,$\mu$Hz.  

There remain conspicuous low-amplitude peaks in the prewhitened periodogram that are adjacent to the $f_2$ and $f_8$ frequencies. These are likely caused by these signals exhibiting slight amplitude or phase variations during the \emph{TESS} observations.  The best-fit frequency values in Table~\ref{tab:freqs} correspond to the highest and central peaks of each mode's power that best represent the intrinsic pulsation frequencies, though the measured amplitudes may be less than the instantaneous maximum amplitudes of these signals during the observations.

\section{Asteroseismic analyses}

As a collaborative effort of the TASC WG8.2, all members with asteroseismic tools and models suited for this data set were invited to contribute their analyses. Two groups submitted full asteroseismic analyses to this effort, which we present in this section. This is the first direct comparison between asteroseismic analyses of the La Plata and Texas groups. By including multiple analyses, we aim to assess the consistency of asteroseismic inferences for pulsating DBVs that utilize different models and methods.

Owing to the quality of the space-based data, the measurements of pulsation mode frequencies presented in Table~\ref{tab:freqs} are reliably accurate and extremely precise. Both analyses that follow aim to interpret this same set of pulsation frequency measurements. 
Certainly the sensitivity of the set of modes detected to the detailed interior structure is a primary limitation on our ability to constrain the properties of this particular DBV.

The combination frequencies are not considered in these analyses, since these are not eigenfrequencies of the star and do not correspond to the pulsation frequencies calculated for stellar models. This highlights the importance of identifying combination frequencies as such; erroneously requiring a model frequency to match a combination frequency would derail any asteroseismic inference.

\subsection{Preliminary mode identification}\label{sec:modeids}

Identifying common patterns in the pulsation spectrum can guide our comparison of the measured frequencies to stellar models. Gravity($g$)-mode pulsations of white dwarfs are non-radial oscillations of spherical harmonic eigenfunctions of the stars. We observe the integrated light from one hemisphere of a star, so geometric cancellation effects \citep{1977AcA....27..203D} typically restrict us to detecting only modes of low spherical degree, $\ell = 1$ or $2$ (modes with one or two nodal lines along the surface). Modes can be excited in a sequence of consecutive radial orders, $k$, for each $\ell$. In the asymptotic limit ($k \gg \ell$), gravity modes of consecutive radial overtone are evenly spaced in period \citep{1990ApJS...72..335T}, following approximately
\begin{equation}\label{eq:asymppers}
{\Pi_{\ell,k}} \approx \Delta \Pi_{\ell}^{\rm a} k + \epsilon  = \frac{{\Pi}_{0}}{\sqrt{\ell(\ell+1)}} k + \epsilon,
\end{equation} 
\noindent where $\Delta \Pi_{\ell}^{\rm a}$ is the period spacing, $\Pi_{0}$ and $\epsilon$ are constants. 

\begin{figure}
	\centering
	\includegraphics[width=1.0\columnwidth]{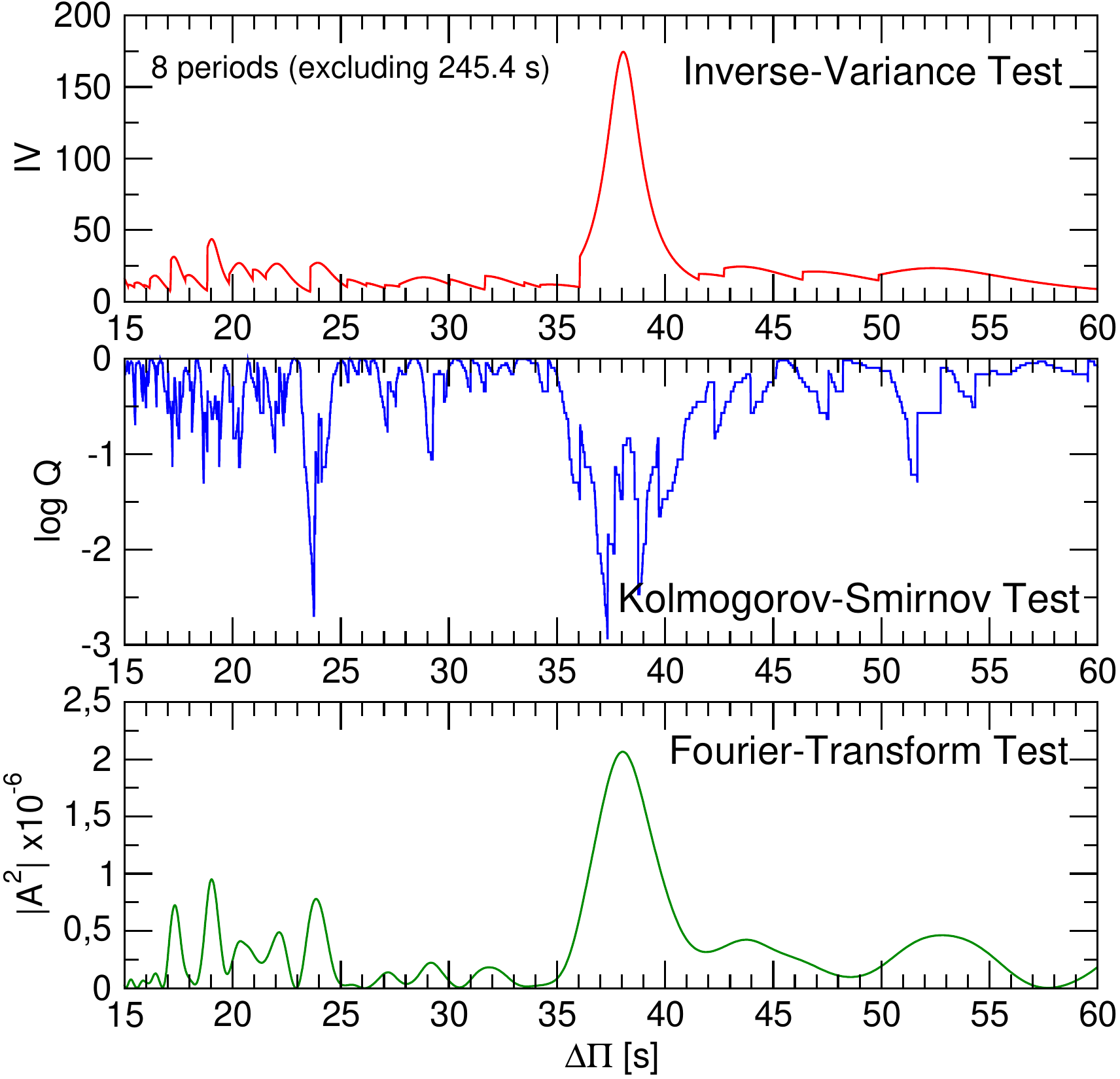}
	\caption{I-V  (upper panel),  K-S (middle panel),  and  F-T (bottom panel) significance  tests  to search for a constant period spacing of TIC\,257459955.  The tests are applied to the pulsation periods in Table~\ref{tab:freqs}, excluding the 245.4\,s period that is not clearly within the asymptotic regime. See text for details.} 
	\label{fig:tests-mass}
\end{figure}

\begin{figure}
	\centering
	\includegraphics[width=1.0\columnwidth]{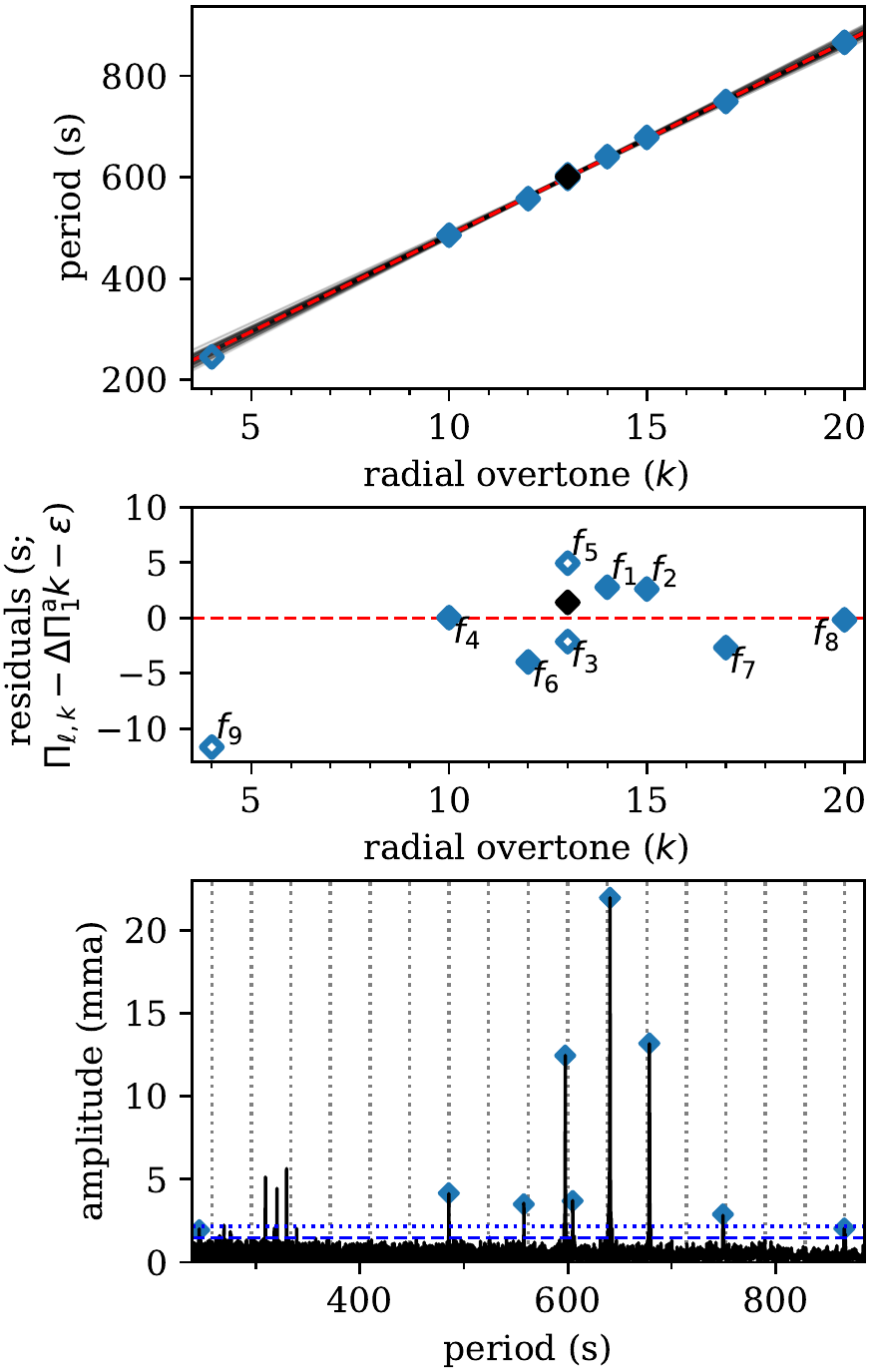}
	\caption{{\sc Top:} The independent pulsation periods of the star plotted versus radial overtone number $k$. The least-squares line fit to $k>5$ (using the mean period of $f_3$ and $f_5$ for $k=13$; black) indicates a roughly constant period spacing consistent with dipole $\ell=1$ modes.  The transparent gray lines represent fits to the perturbed periods assuming different observed azimuthal orders, $m$ (see text). {\sc Middle:} Residuals of the linear fit shows evidence of a possible mode trapping cycle. We label the mode IDs from Table~\ref{tab:freqs}. 
	{\sc Bottom:} The pulsation spectrum in period space with the dotted vertical lines indicating the expected locations of $\ell=1$ modes from the asymptotic pulsation theory given our least-squares fit parameters.}
	\label{fig:l1seq}
\end{figure}

We searched for a constant period spacings in the data of TIC\,257459955 using the  Kolmogorov-Smirnov  
\citep[K-S; see][]{1988IAUS..123..329K}, the inverse variance \citep[I-V; see][]{1994MNRAS.270..222O} and the Fourier Transform \citep[F-T; see][]{1997MNRAS.286..303H} significance tests. 
In the K-S test, the quantity $Q$ is defined as the probability that the observed periods are randomly distributed. Thus, any uniform or at least systematically non-random period spacing in the period spectrum of the star will appear as a 
minimum in $Q$. In the I-V test, a maximum of the inverse variance will indicate a constant period spacing. Finally, in the F-T test, we calculate the Fourier transform of a Dirac comb function (created from a set of observed periods), and then we plot the square of the amplitude of the resulting function in terms of the inverse of the frequency. And once again, a maximum in the square of the amplitude will indicate a constant period spacing. 
In Figure~\ref{fig:tests-mass}  we show the results of applying the tests to the set of periods of Table~\ref{tab:freqs}, excluding the short-period $f_9$ that is not clearly within the asymptotic ($k\gg\ell$) regime.  The three tests
support the existence of a mean period spacing of about 38\,s which
corresponds to our expectations for a dipole ($\ell= 1$) sequence. Note that for $\ell= 2$, according to 
Eq. (\ref{eq:asymppers}), one should find a
spacing of periods of $\sim 22$\,s, which is not observed in our
analysis.\footnote{There is an indication of a $\Delta \Pi \sim 24$ s, that is a bit longer than the prediction for $\ell= 2$ ($\Delta \Pi \sim 22$ s), 
but only from the K-S test.} By averaging
the period spacing derived from the three statistical tests, we found 
$\Delta \Pi= 
38.1 \pm 1.8$\,s as an initial period spacing detection.

We initially obtained a nearly identical result using a frequency solution that did not include $f_8$, as this peak did not exceed our independent significance threshold. Once the preliminary mode identifications were established, it became clear that $f_8$ is located precisely where we expect a $\ell=1$ mode given the asymptotic period spacing.  This prompted us to adopt this mode into our solution for exceeding the lower, frequency-dependent significance threshold, as described in Section~\ref{sec:frequencies}.

This mean period spacing of the $\ell=1$ modes cannot account for the signals at $f_3$ and $f_5$ that are separated by only $19.62\pm0.03\,\mu$Hz ($7.088\pm0.010$\,s). One of these could belong to the quadrupole ($\ell=2$) sequence. Alternatively, $f_3$ and $f_5$ could both be components of a $\ell=1$ rotational multiplet. Stellar rotation causes $2\ell+1$ modes with different azimuthal orders, $m$, to exist for each pair of $\ell$ and $k$ (where $m$ is as integer between $-\ell$ and $\ell$).  These are separated evenly in frequency by an amount proportional to the stellar rotation rate \citep[e.g.,][]{Cox1984}, though many may not be excited to observable amplitude. The frequency separation between $f_3$ and $f_5$ is within the range of rotational splittings of $\ell=1$ modes detected from other pulsating white dwarfs in space-based data \citep[e.g.,][]{Hermes2017}. We leave the exploration of alternate interpretations of these modes up to the individual analyses that follow.

The top panel of Figure~\ref{fig:l1seq} displays a least-squares fit of a line (Eq.\ \ref{eq:asymppers}) through the periods measured for the independent modes listed in Table~\ref{tab:freqs}, given the preliminary period spacing detected from our initial statistical tests. The modes follow a pattern that is consistent with an incomplete $\ell=1$ sequence, though four consecutive modes are detected. 
The absolute radial overtone numbers, $k$, were obtained from the best period-by-period fits from both sets of models described in Sections~\ref{laplata} and \ref{texas}.  We exclude the mode $f_9$ from the fit because its low radial order ($k=4$) is furthest from the asymptotic regime. 
Repeating the fit using alternatively $f_3$ and $f_5$ for the $k=13,\,m=0$ mode has a negligible effect on the best-fit parameters.
The measured periods are weighted equally in the fits since uncertainty in the azimuthal order, $m$, and physical departures from even period spacing likely dominate over the tiny measurement errors\footnote{The error bars on the period measurements are much smaller than the points in Figure~\ref{fig:l1seq}.} in the residuals. 
Using the mean period of $f_3$ and $f_5$ for $k=13$, the best-fit line has $\Delta \Pi_{1}^{\rm a} = 38.1\pm0.3$\,s and $\epsilon = 105\pm5$\,s (Eq.~\ref{eq:asymppers}). This is consistent with the value determined from the three significance tests applied directly to the period list, but the uncertainties are underestimated because they do not account for the $m$ ambiguity. 
We assess our actual uncertainty by repeating fits to 1000 permutations of the periods, each time assigning every observed mode a random $m \in \{-1,0,1\}$ and then correcting to the intrinsic $m=0$ value with an assumed rotational splitting of either $f_3-f_5$ or half that value.  Some representative fits are shaded in the background of Figure~\ref{fig:l1seq}. The standard deviation of best-fit slopes is 0.9\,s, which we add in quadrature to the fit uncertainty for a final measured $\ell=1$ asymptotic period spacing of $\Delta \Pi_{1}^{\rm a} = 38.1\pm1.0$\,s.

The middle panel of Figure~\ref{fig:l1seq} displays the residuals of the measured periods about this fit. We recognize an apparent oscillatory pattern in the residuals 
with a cycle length of $\Delta k\approx 6$, which could correspond to the mode trapping effect of ``sharp'' localized features in the stellar structure \citep[as detected in other DBVs, e.g.,][]{Winget1994}.  These deviations from a strictly even period spacing may provide asteroseismic sensitivity to the location of the helium layer boundary or to chemical composition transitions in the core.
The pulsation spectrum is displayed in units of period in the bottom panel of Figure~\ref{fig:l1seq}, with the expected locations of the $\ell=1$ modes for even period spacing indicated.

\begin{figure}
	\centering
	\includegraphics[width=1.0\columnwidth]{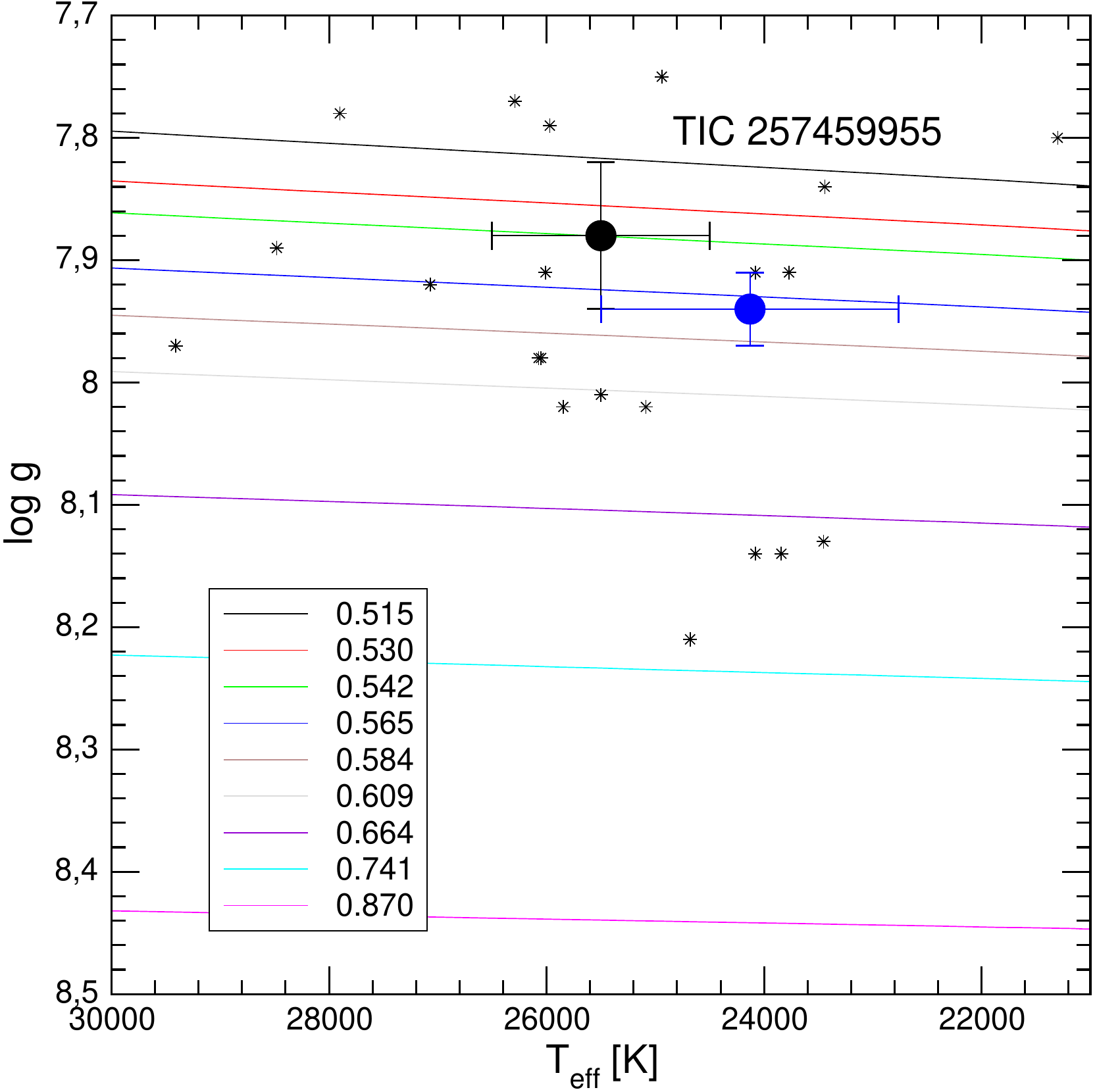}
	\caption{Location of the known DBV stars on the \teff--\logg\ 
	diagram (black star symbols) according to the compilation by 
	\citet{Corsico2019}. The DB white dwarf evolutionary tracks are depicted with  different  colors  according  to  the  stellar  mass. Stellar masses are in solar units. The  location  of TIC 257459955 as given by spectroscopy according to \cite{Voss2007} and \citet{Rolland2018} are highlighted with black and blue circles and error bars. The stellar mass derived from linear interpolation results $M_{\star}= 0.542^{+0.028}_{-0.025} M_{\sun}$ using the data from \cite{Voss2007}, and $M_{\star}= 0.570^{+0.009}_{-0.011} M_{\sun}$ employing the data from \cite{Rolland2018}.} 
	\label{fig:hr-mass}
\end{figure}

\subsection{Analysis from the La Plata group}
\label{laplata}

\begin{figure}
	\centering
	\includegraphics[width=1.0\columnwidth]{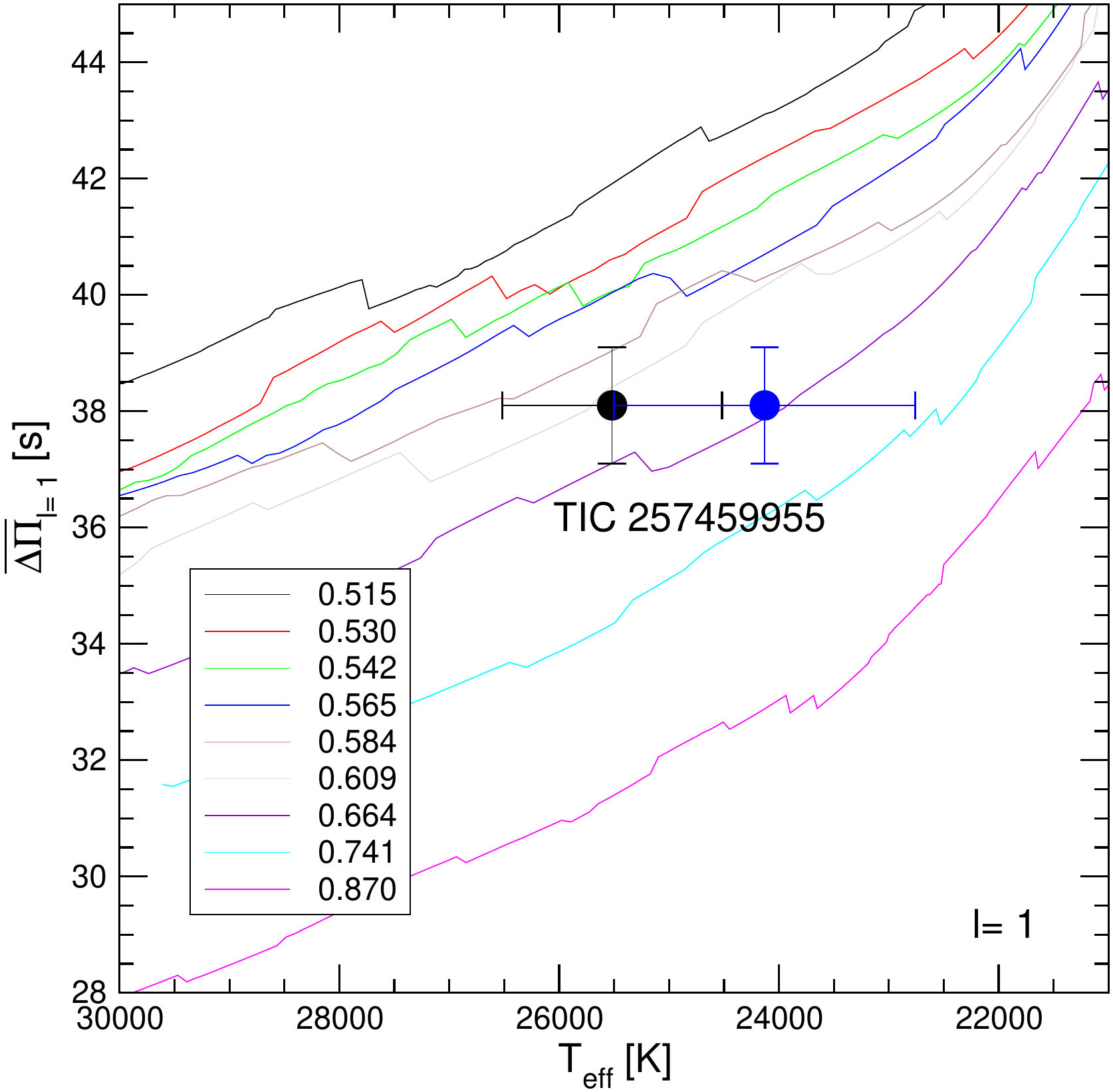}
	\caption{Dipole ($\ell= 1$)  average of the computed   
period spacings, $\overline{\Delta \Pi_{\ell= 1}}$, assessed 
in  a range of periods that includes the periods observed in TIC\,257459955, 
shown as solid curves of different colors for different stellar masses. 
The location of TIC 257459955 when we use the effective temperatures derived by \cite{Voss2007} and \citet{Rolland2018} are highlighted with black and blue circles. We include the 
error bars associated to the uncertainties in $\Delta \Pi$ and $T_{\rm eff}$.
The stellar mass derived from linear interpolation is 
$M_{\star}= 0.621^{+0.057}_{-0.052} M_{\sun}$ ($M_{\star}= 0.658^{+0.106}_{-0.074} M_{\sun}$) by using the $T_{\rm eff}$ derived by 
\cite{Voss2007} \citep{Rolland2018}.} 
	\label{fig:asp-l1-mass}
\end{figure}

In our first analysis, we begin by assessing the stellar mass of TIC\,257459955  
following the methods described in several papers by the La Plata
group on asteroseismic analyses of GW\,Vir stars and DBV stars \citep[see, for instance,][]{2007A&A...475..619C}.  

We first derive the ``spectroscopic'' stellar mass of TIC 257459955 from the $T_{\rm eff}$ and $\log g$ values and appropriate evolutionary tracks. We adopt the values $T_{\rm eff}= 25{,}518 \pm 1000$\,K and $\log g=
7.875 \pm 0.06$ from \cite{Voss2007}\footnote{Note that we adopt uncertainties 
 $\sigma_{T_{\rm eff}}= 1000$ K and $\sigma_{\log g}= 0.06$ as {\it nominal} errors of $T_{\rm eff}$ and and $\log g$.}, and employ the DB white dwarf
evolutionary tracks from \cite{2009ApJ...704.1605A} produced with the 
{\tt LPCODE} evolutionary code. 
These evolutionary 
tracks have been employed in the asteroseismic analyses of the DBV stars KIC\,8626021 
\citep{2012A&A...541A..42C}, KUV 05134+2605 \citep{2014A&A...570A.116B},  
and PG\,1351+489 \citep{2014JCAP...08..054C}. The sequences of DB white dwarf models 
have been obtained taking into account a complete treatment of  
the evolutionary history of progenitors stars, starting from the zero-age main sequence (ZAMS), through the thermally pulsing asymptotic giant branch (TP-AGB) and born-again (VLTP; very late thermal pulse) phases to the domain of the PG\,1159 stars, and finally the DB white dwarf stage. As such, they are characterized by
evolving chemical profiles consistent with the prior evolution. We  
varied the stellar  mass and the effective temperature in our 
model calculations, while the He content, the chemical structure at the CO core, 
and the thickness of the chemical interfaces were fixed by the 
evolutionary history of progenitor objects. 
These employ the ML2 prescription of convection with the mixing length parameter, $\alpha$, fixed to 1 \citep{Bohm1971,1990ApJS...72..335T}.
In Figure \ref{fig:hr-mass} we 
show the evolutionary tracks along with the location  of  all  the  
DBVs known  to  date \citep{Corsico2019}.  We  derive a  new value  of  the  spectroscopic  
mass  for  this  star  on  the  basis  of this  set  of  evolutionary  models.  
This  is  relevant  because  this same  set  of  DB  white dwarf  models  is  used  
below to derive the stellar mass from the period spacing of TIC\,257459955.  
By linear interpolation we obtain an estimate of the spectroscopic mass of 
$M_{\star}= 0.542^{+0.028}_{-0.025} M_{\sun}$ when we use the spectroscopic parameters 
from \cite{Voss2007}, and $M_{\star}= 0.570^{+0.009}_{-0.011} M_{\sun}$ if we adopt the 
spectroscopic parameters from \cite{Rolland2018}. 

In Section~\ref{sec:modeids}, we identified an incomplete dipole ($\ell=1$) sequence of gravity modes with 
high radial order $k$ (long periods) with consecutive modes ($|\Delta  k|= 1$) that are nearly
evenly separated in period by $\Delta \Pi_{1}^{\rm a} = 38.1\pm1.0$\,s. 
This follows our expectations from the asymptotic theory of non-radial stellar pulsations 
given by Eq.~(\ref{eq:asymppers}), where
\begin{equation}
\Pi_0= 2 \pi^2 \left[\int_{r_1}^{r_2} \frac{N}{r} dr\right]^{-1},
\end{equation}
\noindent $N$ being the  Brunt-V\"ais\"al\"a frequency, one of the 
critical frequencies of non-radial stellar pulsations. In principle, 
the asymptotic period spacing or the average of the period spacings computed 
from a grid of models with different masses and effective temperatures can 
be compared with the mean period spacing exhibited by the star to infer the value of the stellar mass. These methods take full 
advantage of the fact that the period spacing of DBV stars primarily 
depends on the stellar mass and the effective temperature, and very weakly 
on the thickness of the He envelope \citep[see, e.g.,][]{1990ApJS...72..335T}. 

We assessed the average period spacings computed for our models
as $\overline{\Delta  \Pi_{\ell=1}}= (n-1)^{-1}  \sum_{k}  \Delta \Pi_k  $,
where the ``forward'' period spacing is defined as
$\Delta \Pi_k= \Pi_{k+1}-\Pi_k$ ($k$ being the radial order) for $\ell=1$ modes
and $n$ is the number of theoretical periods considered from the model.
The theoretical periods were computed with the {\tt LP-PUL} pulsation code \citep{2006A&A...454..863C}. 
For TIC\,257459955, the observed mode periods are $\Pi_k \in [245,866]$\,s. 
In computing the average period spacings for the models, however, we have considered the 
range $[470,1400]$\,s, that is, we excluded short periods that are probably outside the asymptotic regime.
We also adopt a longer upper limit of this range of periods in order 
to better sample the period spacing of modes within the asymptotic regime. 
In Figure~\ref{fig:asp-l1-mass} we  show the  run of  the average of  the
computed period  spacings  ($\ell= 1$) in  terms of  the
effective temperature for our DBV evolutionary sequences, along with 
the observed period spacing for TIC\,257459955.  As can be appreciated from
the Figure, the greater the stellar mass, the smaller the computed values of 
the average period spacing. By means of a
linear interpolation of the theoretical values of
$\overline{\Delta \Pi_{\ell=1}}$,  the measured
$\Delta \Pi$  and spectroscopic effective temperatures
yield  stellar  masses  of
$M_{\star}= 0.621^{+0.057}_{-0.052}  M_{\sun}$ by using the $T_{\rm eff}$ value from 
\cite{Voss2007} and 
$M_{\star}= 0.658^{+0.106}_{-0.074}  M_{\sun}$ by employing the $T_{\rm eff}$ estimate from \cite{Rolland2018}. These stellar-mass values are  
higher than the spectroscopic estimates of the stellar mass.

\begin{figure}
	\centering
	\includegraphics[width=1.0\columnwidth]{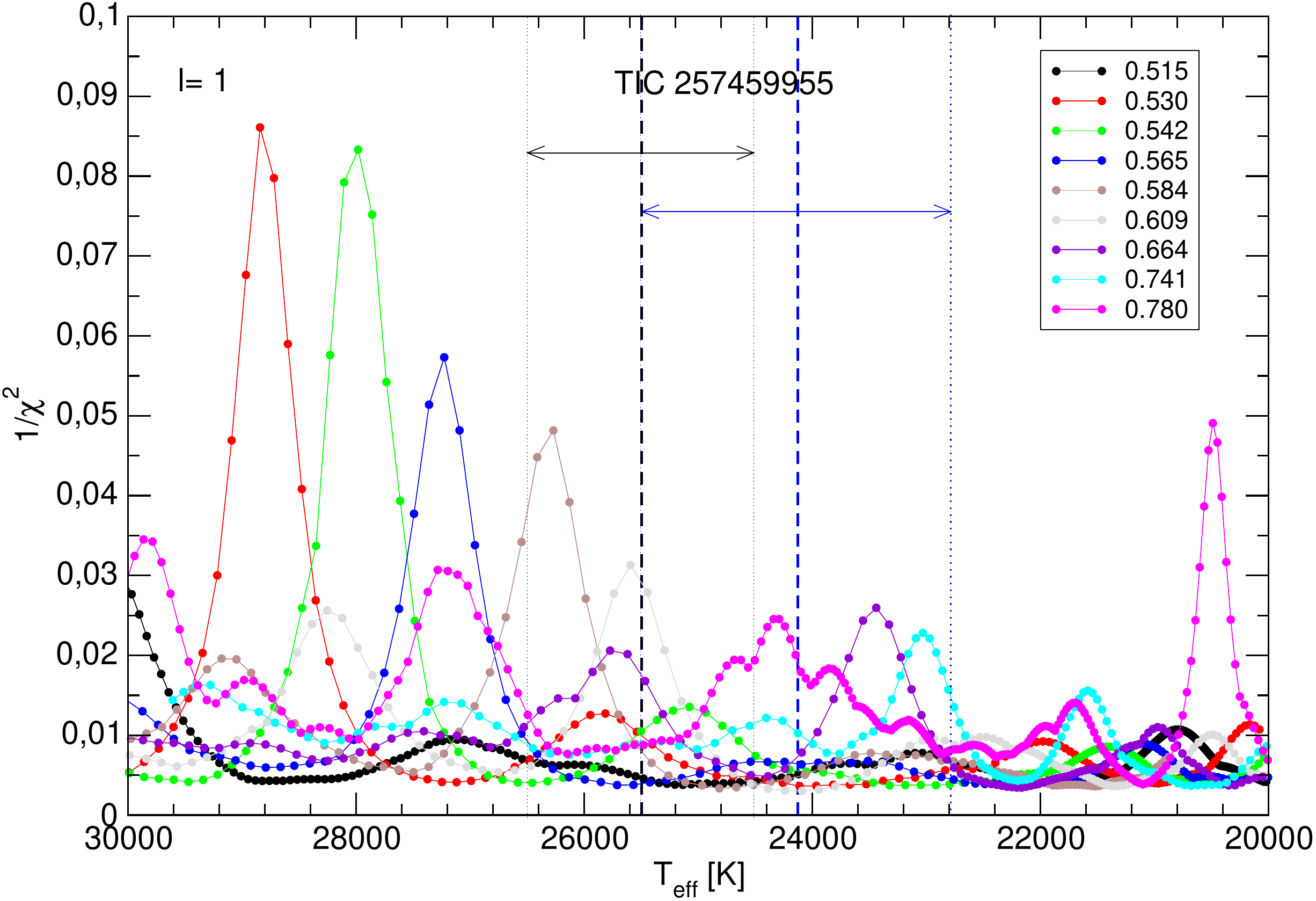}
	\caption{Inverse of the quality function of the period fit in terms of the effective temperature for the case in which we assume that the modes are all $\ell= 1$. The vertical black dashed line indicates the spectroscopic $T_{\rm eff}$ of TIC 257459955 and the vertical dotted lines its uncertainties according to \cite{Voss2007} ($T_{\rm eff}= 25{,}518 \pm 1000$ K). Similarly, the blue vertical lines correspond to the spectroscopic $T_{\rm eff}$ and its uncertainties as derived  by \cite{Rolland2018} ($T_{\rm eff}= 24{,}130 \pm 1369$ K).} 
	\label{fig:chi2-l1}
\end{figure}

\begin{figure}
	\centering
	\includegraphics[width=1.0\columnwidth]{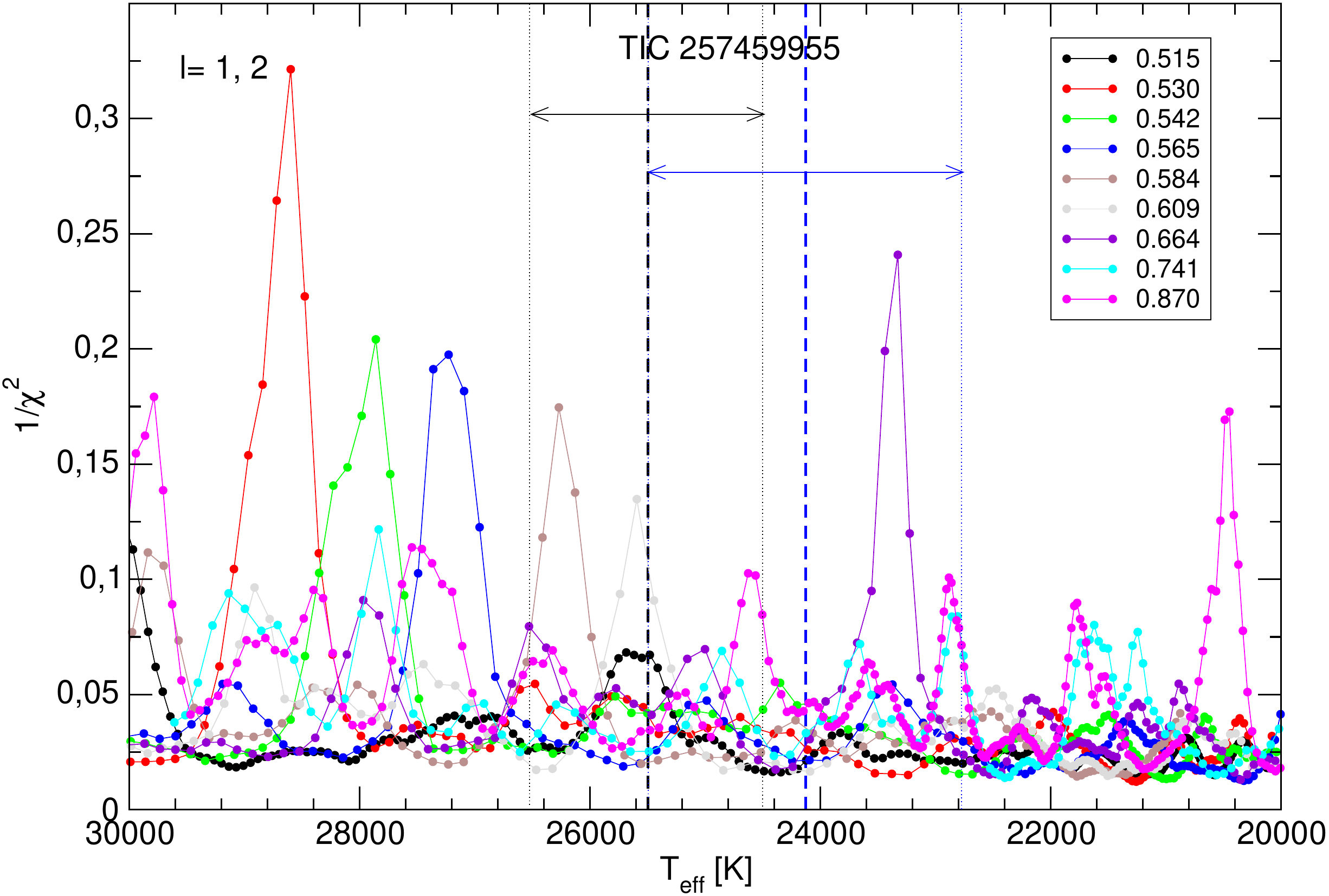}
	\caption{Same as in Figure \ref{fig:chi2-l1}, but for the case in which we assume a mix of $\ell= 1$ and $\ell = 2$ modes.} 
	\label{fig:chi2-l1l2}
\end{figure}

On the other hand, if we instead fix the mass to the value derived from the spectroscopic \logg~ ($0.542$--$0.570\,M_\odot$), then we need to shift the model to higher effective temperature ($\simeq 28,500$~K). This is the result we recover and refine in the period-to-period fitting.
In this procedure we search for a pulsation model that best matches
the \emph{individual} pulsation periods of the star under study. The  
goodness of  the  match  between the   theoretical pulsation  periods
($\Pi_k^{\rm  T}$) and  the observed   individual  periods
($\Pi_i^{\rm  O}$) is assessed by using  a merit function defined
as: 
\begin{equation}
\label{chi}
\chi^2(M_{\star},  T_{\rm   eff})=   \frac{1}{m} \sum_{i=1}^{m}   \min[(\Pi_i^{\rm   O}-   \Pi_k^{\rm  T})^2], 
\end{equation}
\noindent where $m$ is the number of observed periods. The DB white dwarf model that
shows the lowest value of $\chi^2$, if one exists, is adopted as the global ``best-fit
model.''  We assess the function
$\chi^2=\chi^2(M_{\star}, T_{\rm eff})$ for stellar masses in the 
range $[0.515 M_{\sun} -0.741 M_{\sun}]$.  For the effective temperature 
we employ a much finer grid ($\Delta T_{\rm eff} \sim 20$ K) which is given by the time step adopted in the evolutionary calculations of {\tt LPCODE}.
We assumed that the nine pulsation periods of TIC\,257459955 (Table \ref{tab:freqs}) correspond to {\it (i)} modes with $\ell= 1$ only, and {\it (ii)} a mix of $\ell= 1$ and $\ell= 2$ modes. The results are shown in Figs.~\ref{fig:chi2-l1} and \ref{fig:chi2-l1l2}, in which we depict the inverse of the quality function versus $T_{\rm eff}$. Good period fits are associated with maxima in the inverse of the quality function.

Unfortunately, there is no clear and unique solution in the range of effective temperatures from spectroscopy; solutions along the cooling tracks for stellar models with masses $M_{\star}=0.584$--$0.870\,M_{\odot}$ all achieve their best fits to the observed periods at these temperatures. However, global best-fit solutions are found at higher temperature for the stellar model with $M_{\star}= 0.530\,M_{\odot}$, at $T_{\rm eff}= 28{,}844$\,K if all the periods are assumed to be $\ell= 1$ modes and at $T_{\rm eff}= 28{,}600$\,K if the observed periods correspond to a mix of $\ell= 1$ and $\ell= 2$ modes. A good best-fit solution that is in excellent agreement with 
the spectroscopic effective temperature derived by \cite{Voss2007} is found for a model 
with $M_{\star}= 0.609\,M_{\odot}$ and $T_{\rm eff}= 25{,}595$\,K. The chemical profiles and Brunt-V\"ais\"al\"a frequency of the model with $M_{\star}= 0.530 M_{\odot}$ are plotted in Figure~\ref{fig:chemprofs2} in the next section, and the best solution for the $M_{\star}= 0.609\,M_{\odot}$ model that agrees better with the spectroscopic effective temperature is displayed in Figure~\ref{fig:chemprofs3}. If we assume that $f_3$ and $f_5$ are components of a rotational triplet (thus assuming that
they are dipole modes), and consider the average of the 
periods at 597.6\,s and 604.6\,s in our procedure, then the period fits do not improve substantially.

\subsection{Analysis from the Texas group}
\label{texas}

The second asteroseismic fitting analysis that we performed uses models where the chemical profiles are parameterized, along with a few other properties. We used the {\tt WDEC} \citep{Bischoff-Kim18a} with the parameterization of core oxygen profiles described in \citet{Bischoff-Kim18b}. In addition to the 6 core parameters and 5 parameters describing the helium chemical profile, we can also vary the ML2 mixing length coefficient $\alpha$ \citep{Bohm1971} as well as the mass and effective temperature of the model. A $15^{\rm th}$ parameter sets the location of the base of the hydrogen layer, which is not relevant for DBVs.

\begin{figure}
	\centering
	\includegraphics[width=1.0\columnwidth]{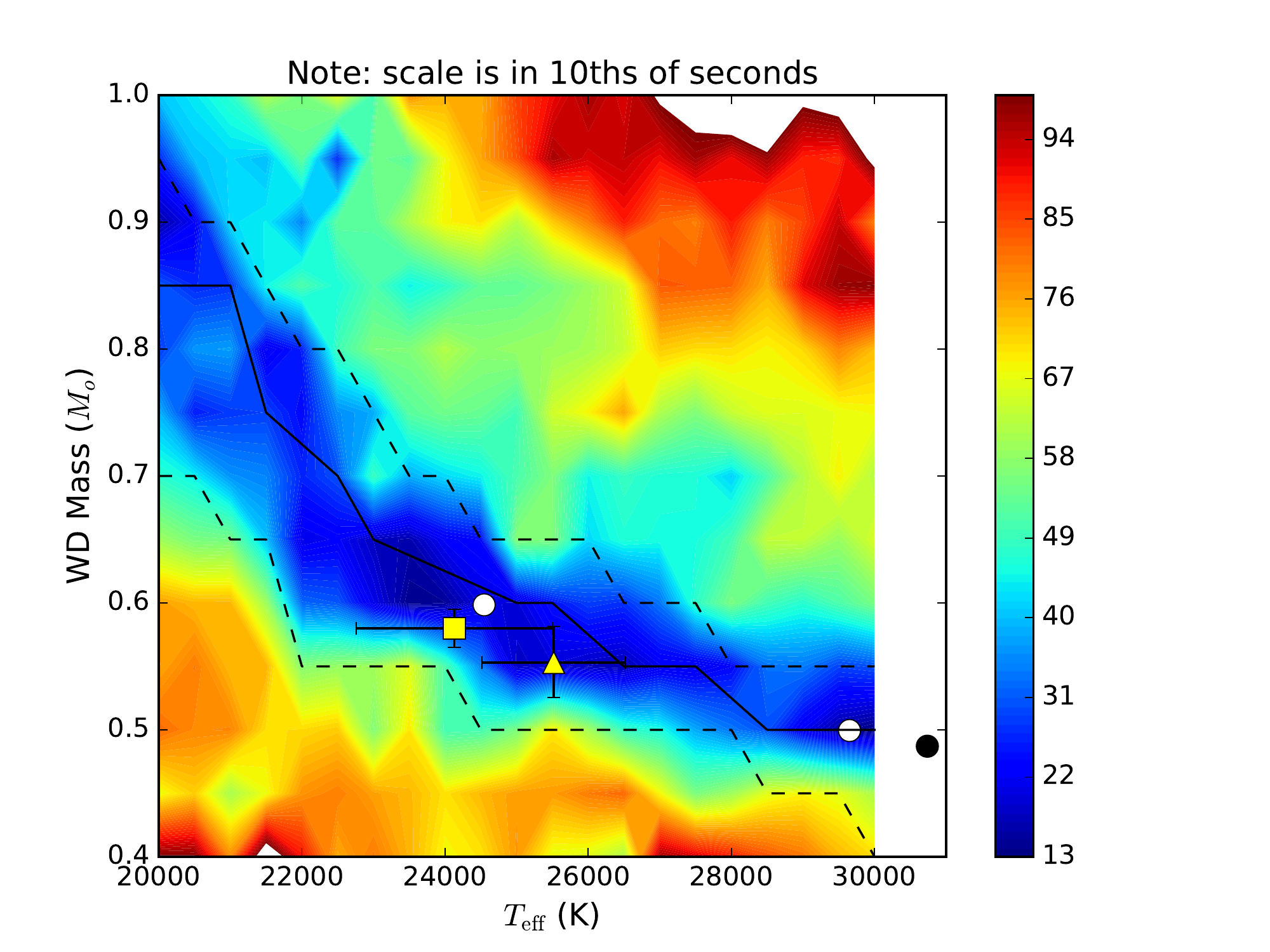}
	\caption{Goodness-of-fit in the mass--\teff\ plane for fit 3 (605~s mode is $m=0$) in units of 10ths of seconds. The triangle is the location of the spectroscopic values from \citet{Voss2007}, and the square is from \citet{Rolland2018}, with error bars indicated. The white and black (colored for visibility) filled circles denote the location of the best fits listed in Table~\ref{tab:bestfitpars}. The solid line is a line of constant period spacing at $\Delta P=38.1$~s and the dashed lines show the 1-second ``error bars'' around that line.} 
	\label{fig:contourplot}
\end{figure}

\begin{table*}
	\centering
	\caption{Parameters of the model grid used in the fits. For a description of each, see \citet{Bischoff-Kim18a} and \citet{Bischoff-Kim18b}. For each parameter, we list the range followed by the step size. $M_{\rm He}$, $M_{\rm env}$, and $M_{\rm H}$ are defined as unitless, negative log fractions of the star by mass.}
	\label{tab:grids}
	\begin{tabular}{lll} 
	\hline
	Oxygen Profile & Helium Profile & Other Grid Parameters\\
	\hline
    $h_1=0,1;0.1$        & $M_{\rm env}=1.5,10;0.5$        & $T_{\rm eff}=20,000-30,000;500$ K \\
    $h_2=0,1;0.1$        & $M_{\rm He}=M_{\rm env},10;0.5$ & $M_{\star}=0.4,1.0;0.05 M_\odot$          \\
    $h_3=0.8$; fixed     & xhe\_bar $=0.3$; fixed          & ML2/$\alpha=0.96$; fixed          \\
    $w_1=0.02,0.52;0.05$ & ${\rm \alpha1}=16$; fixed       & No hydrogen ($M_H$ fixed to 20)   \\
    $w_2=0.15$; fixed    & ${\rm \alpha2}=8$; fixed        &                                   \\
    $w_3=0.36$; fixed    &                                 &                                   \\
    \hline
	\end{tabular}
\end{table*}

\begin{table*}
	\centering
	\caption{Best fit parameters, result of simplex search.}
	\label{tab:bestfitpars}
	\begin{tabular}{cccccccccc} 
		\hline
		fit ($m=0$ mode)  & \multicolumn{2}{c}{Stellar parameters} & \multicolumn{2}{c}{Envelope parameters} & \multicolumn{3}{c}{Core parameters} & $1/\chi^2 \; (1/s^2)$ & $\sigma_{\rm RMS} (s)$ \\
		                  & \multicolumn{2}{c}{$T_{\rm eff}, M_{\star}$} & \multicolumn{2}{c}{$M_{\rm env}, M_{\rm He}$} & \multicolumn{3}{c}{$h_1,h_2,w_1$}  & & \\
		\hline
		1 (598~s)   & 30737~K & $0.487 M_\odot$ & 1.505 & 6.158 &  0.721 & 0.246 & 0.370 & 14.2 & 0.283    \\ 
		2 (601~s)   & 24546~K & $0.598 M_\odot$ & 1.527 & 4.500 &  0.459 & 0.405 & 0.595  & 2.47 & 0.680   \\
		3 (605~s)   & 29650~K & $0.499 M_\odot$ & 3.595 & 6.411 &  0.595 & 0.123 & 0.375 & 14.7 & 0.279  \\ 
		\hline
	\end{tabular}
\end{table*}

\begin{table*}
	\centering
	\caption{List of periods for the best fit models. All modes are $\ell=1$.}
	\label{tab:periodfit}
	\begin{tabular}{cccccccc} 
		\hline
		mode    & k  & observed period & fit 1     & fit 2    & fit 3    \\
		ID      &    & (s)             & (s)       & (s)      &   (s)    \\
		\hline
		$f_9$   & 4  & 245.4399        & 245.5704  & 245.7442 & 245.4326 \\
		$f_4$   & 10 & 485.5275        & 485.7931  & 486.3560 &  485.5028 \\
		$f_6$   & 12 & 557.6493        & 557.3893  & 557.1444 &  558.0835 \\
		$f_3$   & 13 & 597.5538        & 597.9700  &          &          \\
		Inferred& 13 & 601.1           &           & 602.0    &          \\
		$f_5$   & 13 & 604.6423        &           &          & 604.1388 \\
		$f_1$   & 14 & 640.5330        & 640.5687  & 641.1684 & 640.7684 \\
		$f_2$   & 15 & 678.4328        & 678.4701  & 678.0488 & 678.2318 \\
		$f_7$   & 17 & 749.2676        & 748.9701  & 748.9309 & 749.3228 \\
		$f_8$   & 20 & 865.9744        & 865.5980  & 865.1238 & 865.9226 \\
		\hline
		$\sigma_{\rm RMS}$ (s)  &      &    & 0.283    & 0.680  & 0.279   
	\end{tabular}
\end{table*}

We had to fix some parameters in order to keep the problem computationally tractable and also constrained. We fixed ML2/$\alpha$ to 0.96 (see \citealt{Bischoff-Kim18a}) and some oxygen and helium profile parameters to values such that we reproduced profiles from \citet{Dehner95} and \citet{2009ApJ...704.1605A}. 
\citet{Bischoff-Kim2015} demonstrated that varying the mixing length parameter has a negligible effect on the pulsation periods in the range observed for TIC\,257459955.
We did allow three of the oxygen profile parameters ($h_1$, $h_2$, and $w_1$) to vary, as well as two of the helium profile parameters (the location of the base of the helium envelope $M_{\rm env}$ and the pure helium layer mass $M_{\rm He}$).\footnote{For shorthand, these helium profile parameters are defined as the negative log fractions of the star by mass; e.g., $M_{\rm env} = 1.5$ means, in mass units, $M_{\rm env} = 10^{-1.5}\times M_{\star}$.}
In addition, we varied the mass and the effective temperature of the models, for a total of 7 parameters. These parameters were determined to be the ones that had the greatest effect on the quality of the fits.

We started our model comparison with a grid search to locate minima in the global parameter space. The values of the parameters calculated in our grid are listed in Table~\ref{tab:grids}.
In our comparison of the measured periods to the models, we considered three different values for the $m=0$ component in the 598, 605~s multiplet: we tested each of these as the central component individually, as well as their average, which could be undetected between two observed $m=\pm1$ modes. 
We refer to our results from different assumptions of the $m=0$ component of this mode as fits 1--3 in order of increasing period.

We finish with a simplex search \citep{simplex} to refine the minima, calculating {\tt WDEC} models on the fly as the algorithm sought to minimize $\sigma_{\rm RMS}$. The simplex method explores the parameter space of the models on its own; it does not reference the models from our initial grid, nor is it bound to the same mass and effective temperature limits. In fact, some of the best-fit models returned are hotter than 30,000\,K.  
We list the parameters of the best fit models in Table~\ref{tab:bestfitpars}, along with a measure of the quality of fit, computed the same way as in Section~\ref{laplata} (Eq.~\ref{chi}) to facilitate comparison with the previous analysis. We also list a different measure of goodness of fit, the standard deviation $\sigma_{\rm RMS}$, because that is the quantity minimized in our grid and simplex searches:
\begin{equation}
     \sigma_{\rm RMS} = \sqrt{\frac{m}{m-1}*\chi^2}
\end{equation}
\noindent where again, $m$ is the number of observed periods. We list the periods of the best fit models in Table~\ref{tab:periodfit}. 

We show the best fit contour map for fit 3 (605~s mode is $m=0$) in the mass vs.\ effective temperature parameter plane in Figure~\ref{fig:contourplot}, along with the location of the best fits and lines of constant period spacing corresponding to the value derived from the period spectrum of the star. The contour plots for other $m=0$ choices look similar. The period spacing for the models is calculated by fitting a line through the higher $k$ modes ($k=11$ and up) and determining the slope, in the same way we use the linear fit of Figure~\ref{fig:l1seq} to determine one value for the average period spacing present in the pulsation spectrum of the star. The limit of $k=11$ was chosen by visual inspection. Modes of higher radial overtone follow a linear trend closely, and so are reflective of the asymptotic period spacing discussed in Section \ref{sec:modeids}. The computation of period spacings for the models in the grid are further discussed in \citet{Bischoff-Kim2019}. The correlation between the quality of fits and period spacing is striking. This is to be expected for this object with such a tight linear sequence of (assumed) $\ell=1$ modes (Figure~\ref{fig:l1seq}).

To determine the location of the spectroscopic points in Figure~\ref{fig:contourplot}, we interpolated WDEC models to translate the \logg\ measurements for the star into mass. We find $M_{\star}=0.553^{+0.029}_{-0.027} M_\odot$ for the \citet{Voss2007} measurements and $M_{\star}=0.580^{+0.014}_{-0.015} M_\odot$ for the \citet{Rolland2018} values.  The slight difference compared to the values inferred by the La Plata models comes from the fact that surface gravity depends not only on the mass and effective temperature, but also on the interior structure of the models, and we use different models. 

\begin{figure}
	\centering
	\includegraphics[width=1.0\columnwidth]{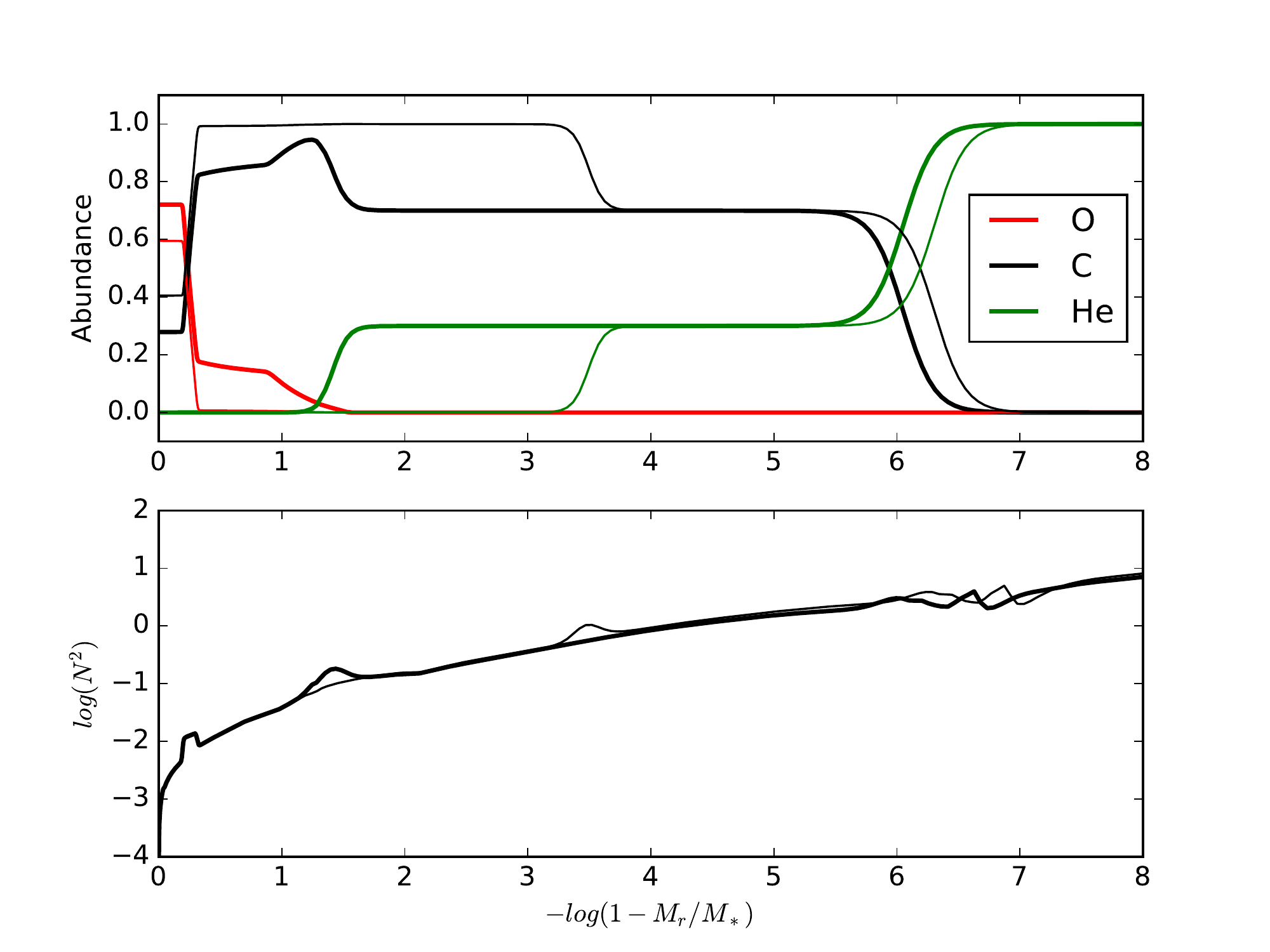}
	\caption{Top panel: chemical abundance profiles for the two best fit models of Table \ref{tab:bestfitpars} (fits 1 and 3 with higher effective temperatures). The bold lines correspond to fit 1, while the thin lines correspond to fit 3. Bottom panel: the corresponding Brunt-V{\"a}i{\"a}sal{\"a} frequency curves.} 
	\label{fig:chemprofs1}
\end{figure}

The two best global fits from the simplex search by a significant margin are fit 1 and fit 3. Both are at high effective temperature, inconsistent with the spectroscopic value. They differ mainly by the thickness of the helium envelope, $M_\mathrm{env}$. We compare their chemical abundance profiles and Brunt-V{\"a}i{\"a}sal{\"a} frequency curves in Figure~\ref{fig:chemprofs1}.
We note a possible manifestation of the core-envelope symmetry here, as has been observed in the asteroseismic fitting of the DBV GD\,358 and discussed in \citet{Montgomery03}. The two models differ in the location of bumps in their Brunt-V{\"a}i{\"a}sal{\"a} frequencies corresponding to the transitions from pure carbon to a mix of carbon and helium (at $\log(1-M_r/M_{\star})\simeq 1.5$ and $\log(1-M_r/M_{\star})\simeq 3.5$). The bumps have similar shapes. In the core-envelope symmetry, a feature in the core (or in this case deep in the envelope) can be replaced by a feature further out and produce a similar period spectrum. This will result in two models that fit almost equally well, or in this case a significant change in the location of a Brunt-V{\"a}i{\"a}sal{\"a} feature between best-fit models that use slightly different periods for the $k=13$ mode. The central oxygen abundance and the transition from a mix of helium and carbon to pure helium have a weaker effect on the periods.

\begin{figure}
	\centering
	\includegraphics[width=1.0\columnwidth]{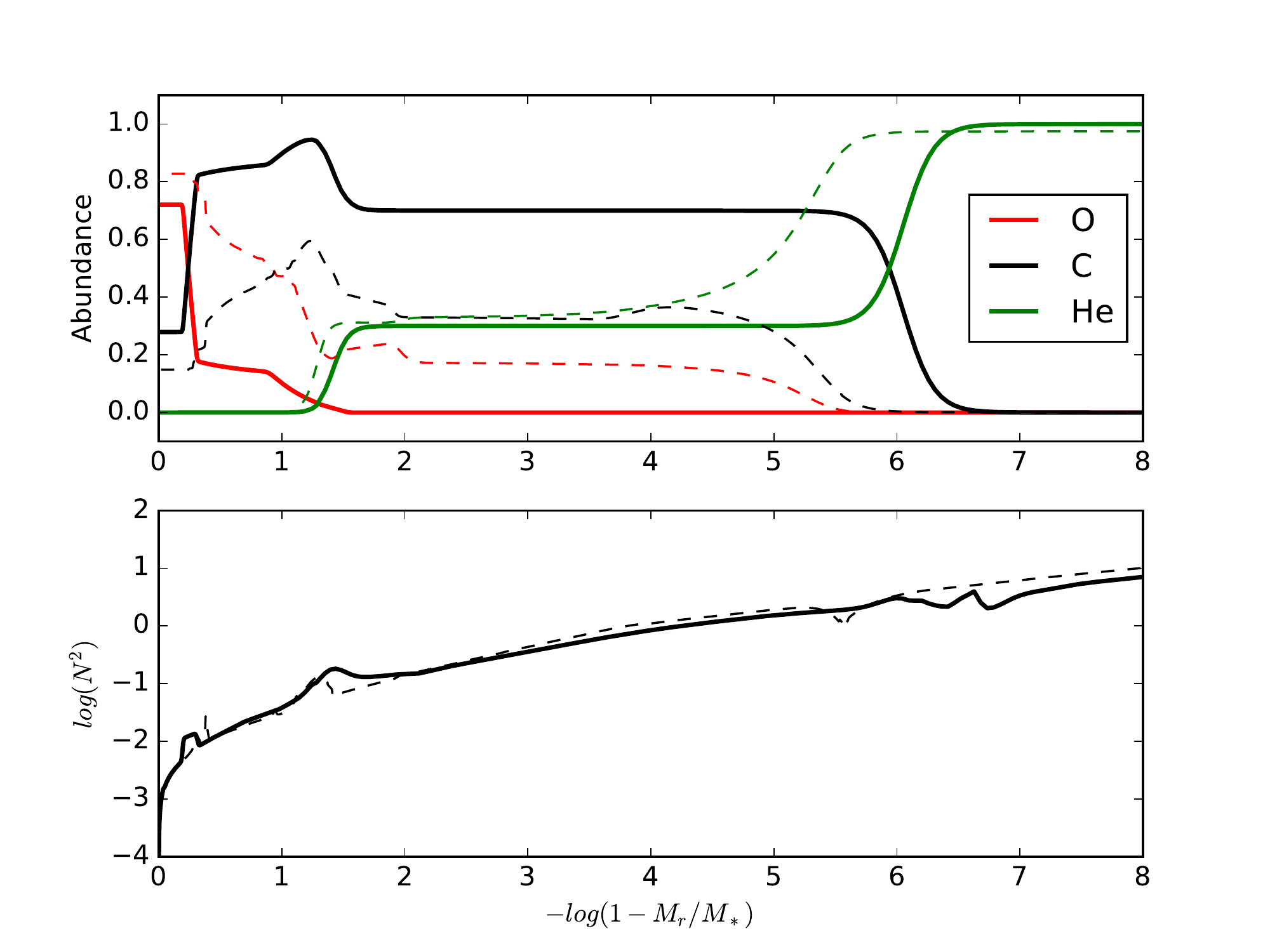}
	\caption{Top panel: chemical abundance profiles for fit 1 of Table \ref{tab:bestfitpars} (solid lines), as well as the best fit model of Section~\ref{laplata} (dashed lines). We chose to contrast these two best fit models because of their similarities. Bottom panel: the corresponding Brunt-V{\"a}i{\"a}sal{\"a} frequency curves.}
	\label{fig:chemprofs2}
\end{figure}

Fit 1 agrees closely with the La Plata model that we found to produce the best global period-by-period fit in Section~\ref{laplata}.
Considering that we base our fixed parameter values on models from Section~\ref{laplata}, that is expected.
We compare the chemical and Brunt-V{\"a}i{\"a}sal{\"a} profiles of fit 1 with the best-fit La Plata model in Figure~\ref{fig:chemprofs2}.
These are similar in the location of the chemical transition zones, which are mainly responsible for setting the period spectrum of a model, so this is not by accident.

\begin{figure}
	\centering
	\includegraphics[width=1.0\columnwidth]{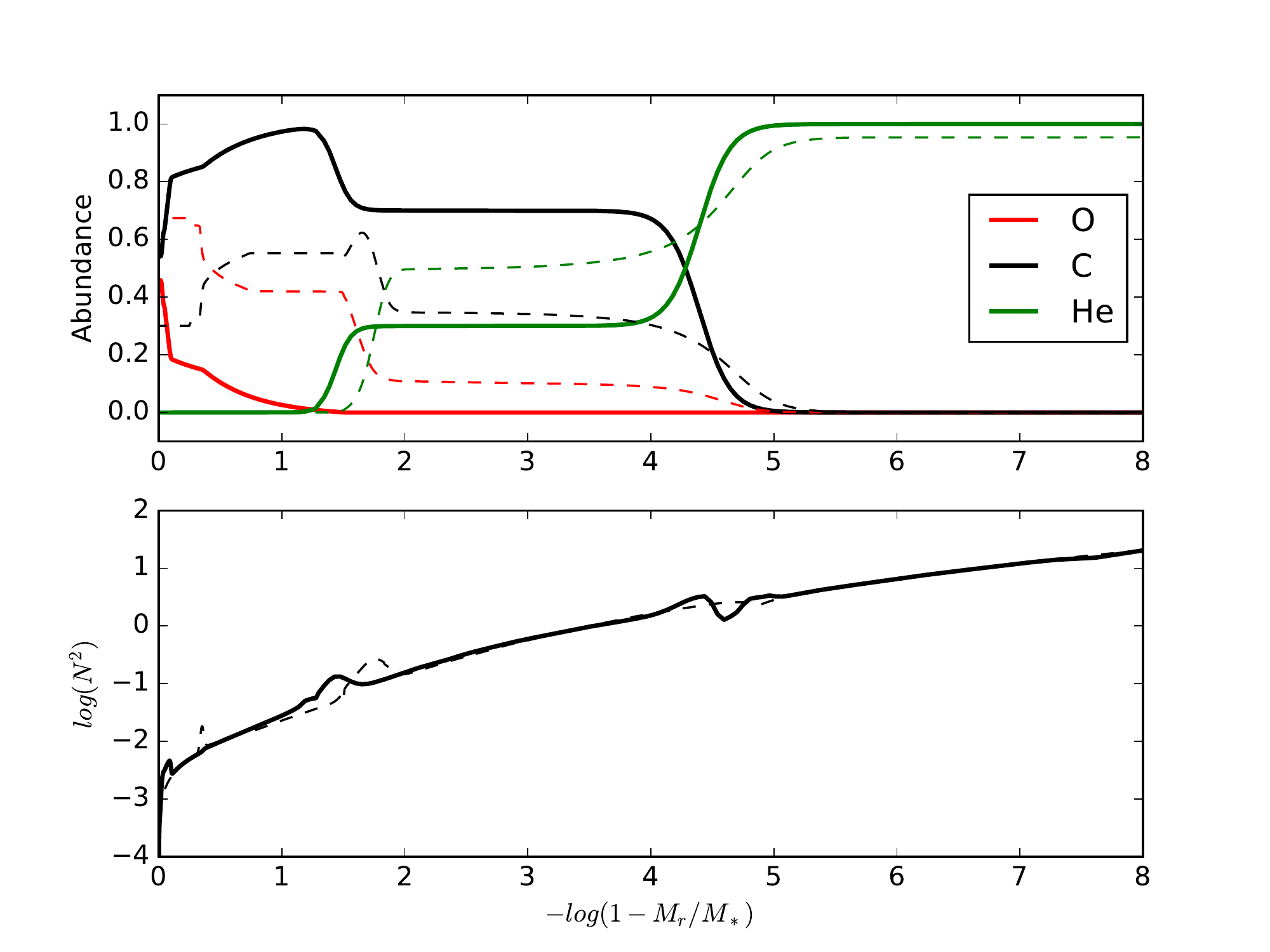}
	\caption{Top panel: chemical abundance profiles for fit 2 of Table \ref{tab:bestfitpars} (solid lines), as well as the best fit model for the 0.609\,\msun\ evolutionary track from Section~\ref{laplata} (dashed lines). These secondary solutions are in better agreement with constraints from spectroscopy. Bottom panel: the corresponding Brunt-V{\"a}i{\"a}sal{\"a} frequency curves.}
	\label{fig:chemprofs3}
\end{figure}

In addition, we also have one good fit at lower effective temperature using the average of 598 and 605\,s periods for $m=0$ (fit 2). 
While this is not nearly as good of a fit to the observed periods, it does agree within uncertainties of both spectroscopic measurements of WD\,0158$-$160's effective temperature.
In Figure~\ref{fig:chemprofs3} we compare the chemical profiles and Brunt-V{\"a}i{\"a}sal{\"a} frequency of this fit to the best-fit solution along the 0.609\,\msun\ evolutionary track from the La Plata models (Section~\ref{laplata}), which also agrees with the spectroscopic \teff. We see again that the features that most affect the period spectrum in the profiles of these cooler secondary solutions appear at roughly the same locations.

As mentioned in Section~\ref{sec:modeids}, the 598~s/605~s doublet is consistent with a rotationally split $\ell=1$ mode. The rotation frequency $\Omega$ of the white dwarf is related to the frequency splitting $\Delta \sigma$ by a relation that involves the $m$ identification of the mode and a mode-dependent factor $C_{kl}$ \citep{Unno89}:
\begin{equation}
    \Delta \sigma = m(1-C_{kl})\Omega
\end{equation}

For our best-fit models we find that the 598~s mode has $C_{kl}=0.49$. Using that value and assuming one of the members of the doublet is the $m=0$ mode, we find a rotation period of 7\,hrs. If we have instead observed the $m=1$ and $m=-1$ components of the triplet, then the rotation period is 14\,hrs. Both are consistent with the rotation periods expected empirically for white dwarf stars \citep{Kawaler15,Corsico2019}.

\section{Discussion and conclusions}

\emph{TESS} observed the pulsating helium-atmosphere DBV white dwarf WD\,0158$-$160 as TIC\,257459955 for 20.3 nearly uninterrupted days in Sector 3 at the short 2-minute cadence.  These data enabled accurate determination of the pulsation frequencies to $\sim0.01\,\mu$Hz precision.  Our frequency analysis reveals nine significant independent pulsation modes and eleven combination frequencies.  The pattern of the observed pulsations is consistent with an incomplete sequence of dipole $\ell=1$ modes with an asymptotic mean period spacing of $38.1\pm1.0$\,s. Two modes separated by 19.6\,$\mu$Hz could belong to a rotationally split $\ell=1$ triplet, implying a stellar rotation period of 7 or 14 hours, depending on which components are being observed.

The shortest-period pulsation at 245\,s was included in our frequency solution based on corroboration with archival photometry from \citet{Kilkenny2016}. It appears that a different set of modes were dominant in those ground-based observations that first revealed WD\,0158$-$160 to be a DBV pulsator. Seasonal changes such as these have been observed in other DBVs such as GD\,358 \citep[e.g.,][]{Bischoff-Kim2019}. The slight residuals in the periodogram of the fully prewhitened time series (Figure~\ref{fig:ft}) near modes $f_2$ and $f_8$ likely indicate that these modes were varying in amplitude during the \emph{TESS} observations.

Enabled by recent improvements in the {\tt WDEC} \citep{Bischoff-Kim18a} and fostered by the collaborative TASC WG8.2, we present for the first time a direct comparison between asteroseismic analyses from the La Plata and Texas groups. A primary difference between the two sets of models is that the La Plata group uses fully evolutionary models calculated with {\tt LPCODE}, while the Texas group computes grids of structural models with parameters sampled on demand using {\tt WDEC}. Both groups find that the measured mean period spacing of $\ell=1$ modes traces paths of good model agreement of decreasing mass with increasing effective temperature (Figs.~\ref{fig:asp-l1-mass} and \ref{fig:contourplot}) that pass through the average DB white dwarf mass of $\approx0.62$\,\msun\ \citep[e.g.,][]{Kepler2019} at the spectroscopic effective temperature of $25{,}500\pm1000$\,K from \citet{Voss2007}.

When considering individual mode periods, both analyses achieve excellent asteroseismic fits to models with $T_\mathrm{eff}$ in excess of $\sim 28{,}500$\,K and lower masses $M_\star\approx$ 0.5\,\msun. These solutions are significantly hotter than the spectroscopic effective temperatures 
obtained by \citet{Voss2007} and \citet{Rolland2018}.
External uncertainties in \teff\ measured from spectroscopy can be as high as $\approx1000$\,K for DB white dwarfs in the DBV instability strip \citep[e.g.,][as assumed in Section~\ref{laplata}]{Beauchamp1999}, but this is insufficient to bring our optimal seismic fits into agreement with the spectroscopic values.
Corrections for spectroscopically determined DB atmospheric parameters based on 3D convection simulations also cannot account for this discrepancy \citep{Cukanovaite2018}. 
The only DBV reported to have $T_\mathrm{eff} > 30{,}000$\,K is PG\,0112$+$104, with spectroscopic parameters $T_\mathrm{eff} = 31{,}300\pm500$\,K and $M_\star = 0.52\pm0.05\,M_\odot$ \citep{Dufour2010} and variability dominated by shorter-period ($\sim200$\,s) pulsations \citep{Hermes2017a}.

Both analyses also yield good fits as secondary solutions that are well in line with the spectroscopic measurements.
The La Plata evolutionary track for a 0.609\,\msun\ DB achieves its best fit at $T_\mathrm{eff}=25{,}595$\,K, and WDEC model 2 (assuming 601.1\,s for the $k=13$ mode) has $T_\mathrm{eff} = 24{,}546$\,K and a seismic mass of 0.598\,\msun.

The structural profiles of the global best-fit models from both analyses are compared in Figure~\ref{fig:chemprofs2}, and Figure~\ref{fig:chemprofs3} shows the same for the secondary solutions that agree with spectroscopy. While at first glance, it might appear that there is little agreement between the models from our two analyses, it is important to note that the transition zones (and the corresponding features in the Brunt-V{\"a}i{\"a}sal{\"a} frequency) do approximately line up. It is well known that pulsation periods are most sensitive to the location of the features in the Brunt-V{\"a}i{\"a}sal{\"a} frequency, and less so to their shape \citep[e.g.,][]{Montgomery03}. The consistency of the results of the two seismic analyses is encouraging, as it supports that the results are not dominated by extrinsic errors from the choice of models. Conducting these two analyses in parallel also helps us to select preferred fits for this star.

We can convert the luminosities of our best-fitting 
models into seismic distances for comparison with the precise astrometric distances available from \emph{Gaia} DR2 \citep{Gaia2018}.
Model 2 from Table~\ref{tab:bestfitpars} is the {\tt WDEC} solution with the lowest luminosity $\log L/L_{\odot} = -1.258$ and a bolometric correction \citep{Koester2018} of ${\rm B.C.}=-2.55$\,mag.
We use the well-known formulas 
$M_\mathrm{Bol} = M_{\mathrm{Bol},{\odot}} -2.5 \log (L/L_{\odot})$ and $M_V = M_\mathrm{Bol}-{\rm B.C.}$ to solve for the absolute visual magnitude of $M_V=10.44$ for this model, where $M_{\rm Bol}$ is its absolute bolometric magnitude, and $M_{\mathrm{Bol},{\odot}} = 4.74$.
The apparent visual magnitude of TIC\,257459955 from the Fourth US Naval Observatory CCD Astrograph Catalog \citep{Zacharias2012} is $m_V =14.55\pm 0.08$, which agrees with the \emph{Gaia} DR2 magnitude of $14.53\pm0.01$ in the similar $G_\mathrm{BP}$ passband \citep{Gaia2018}.
Applying the formula $5\log d = m_V + 5 - M_V$, we find that the {\tt WDEC} model 2 visual magnitude scales to the observed apparent magnitude at a seismic distance of $64.0$--$68.9$\,pc, 
which agrees with the \emph{Gaia} DR2 distance of $68.14 \pm 0.28$\,pc from \citet{BailerJones2018} within the error bars. 
The secondary 0.609\,\msun\ La Plata solution within the spectroscopic temperature range similarly agrees with the \emph{Gaia} distance constraint.  
However, hotter global solutions from both {\tt WDEC} (models 1 and 3) and {\tt LPCODE} have temperatures $T_{\rm eff} \approx 30{,}750$--$28,800$\,K; the luminosity range is 
$\log L/L_{\odot} = -0.708$ to $-0.867$, which implies a bolometric correction of $-3.22$ to $-3.06$ magnitudes. Using the above formulas, these models yield asteroseismic distances $79.8$--$95.5$\,pc, which are much further than the distance to TIC\,257459955 from \emph{Gaia} DR2. 

Because of their disagreement with the spectroscopic $T_\mathrm{eff}$ and the parallax, we regard {\tt WDEC} models 1 and 3 and the global best-fit {\tt LPCODE} model as less 
likely solutions, in spite of their excellent fits to the pulsation periods. We prefer the cooler, secondary solutions from both sets of models that agree with both astrometry and spectroscopy as better representations of TIC\,257459955, and we conclude that these external constraints are necessary for selecting the best seismic model given the sensitivity of this particular observed set of modes to the interior stellar structure.

\begin{table}
	\centering
	\caption{\emph{Gaia} data for the common proper motion stars.}
	\label{tab:Gaia}
	\begin{tabular}{lcc} 
		\hline
		Parameter & TIC\,257459955 & G272-B2B  \\
		\hline
		RA  (deg) & $02{\rm h}00{\rm m}56.8502{\rm s}$  &  $02{\rm h}00{\rm m}56.9138{\rm s}$ \\
		Dec (deg) &  $-15^\circ 46\arcmin 09.2467\arcsec$ &  $-15^\circ 46\arcmin 16.997\arcsec$ \\
		$G$ (mag) &  $14.6789\pm0.0018$   & $14.8075\pm0.0006$ \\
		$G_\mathrm{BP}$ (mag) &  $14.527\pm0.011$  & $16.238\pm0.010$  \\
		$G_\mathrm{RP}$ (mag) &  $14.840\pm0.008$  & $13.631\pm0.002$  \\
		parallax (mas) & $14.64\pm0.06$ & $14.55\pm0.07$ \\
		RA pm (mas\,yr$^{-1}$) & $127.47\pm0.10$  & $125.84\pm0.12$ \\
		Dec pm (mas\,yr$^{-1}$) & $31.14\pm0.08$  & $ 29.12\pm0.10$ \\
		distance\tablefootnote{From \citet{BailerJones2018}.} (pc) & $68.14 \pm 0.28$ & $68.58 \pm 0.33$\\ 
		\hline
	\end{tabular}
\end{table}

The possibility that the nearby red dwarf G272-B2B is a common proper motion companion to WD\,0158$-$160, as discussed by \citet{Kilkenny2016}, is also supported by the \emph{Gaia} DR2 astrometric data for these stars, which we summarize in Table~\ref{tab:Gaia}.
At a distance of $68.58 \pm 0.33$\,pc \citep{BailerJones2018}, G272-B2B has $M_V = 9.29$, and using Table~$15.7$ in \citet{Drilling2000}, this absolute magnitude is consistent with 
a type M1 dwarf, which is a bit more luminous than suggested by \citeauthor{Kilkenny2016}. With an on-sky separation of only 7\arcsec\ compared to the \emph{TESS} plate scale of 21\arcsec\,pix$^{-1}$, G272-B2B will contribute significant light to the photometric aperture of TIC\,257459955. In fact, the header keyword {\tt CROWDSAP} from the PDC pipeline suggests that only 30\% of the total flux originally measured in the aperture is from the white dwarf target WD\,0158$-$160, which has the effect of decreasing the signal-to-noise of the periodogram by a factor of 1.8, potentially obscuring lower-amplitude pulsation signals.\footnote{The PDC pipeline subtracts off the expected contributions from sources other than the target so that amplitudes measured for detected pulsations should be accurate \citep{Twicken2010}.}

We note a striking similarity between the pattern of pulsation modes observed in TIC\,257459955 and the prototypical DBV variable GD\,358 \citep{Winget1982}. Over three decades of observations have revealed a clear pattern of nearly sequential $\ell=1$ modes in GD\,358 with a mean period spacing of 39.9\,s \citep{Bischoff-Kim2019}.  This is similar to the period spacing measured from the \emph{TESS} observations of TIC\,257459955, but the periods of the corresponding modes in GD\,358 are all longer by $\approx20$\,s. A comparative seismic analysis of these stars could reveal how this relative translation of mode periods results directly from small differentials in their physical stellar parameters.

The pulsation frequencies calculated for stellar models have azimuthal order $m=0$, corresponding to the central components of rotationally split multiplets.  Generally, not all components of a multiplet are detected in pulsating white dwarfs, and the observed modes are simply assumed to be $m=0$ in the absence of other information.  This can introduce discrete inaccuracies in the fitting of each period of a few seconds, compared to the millisecond precision that these periods are measured to from \emph{TESS} photometry. The analysis from the Texas group (Section~\ref{texas}) demonstrated the non-negligible effect of this uncertainty for just a single radial order on the inferred stellar structure, treating each of $f_3$, $f_5$, and their average as the $m=0$ mode (Table~\ref{tab:bestfitpars}). 
As a manifestation of a core-envelope symmetry, these small changes to a single mode period resulted in significant differences in the best-fit location of the base of the helium envelope in otherwise similar models, as displayed for fits 1 and 3 in Figure~\ref{fig:chemprofs1}.
\citet{Metcalfe2003} argued from Monte Carlo tests that fitting models with the assumption of $m=0$ for modes detected from ground-based observations of DBVs yields the same families of solutions and often the same best-fit model (with root-mean-square period differences of $\approx1$\,s) as when reliable $m$ identifications are available; however, we should be wary of whether these results hold in the era of space photometry, as model fits are now being achieved to the unprecedented precision of our current period measurements \citep{Giammichele2018}. 

Our interpretation of the detected signals from the \emph{TESS} data did not consider their observed amplitudes. We identified the majority of peaks detected in the periodogram as nonlinear combination frequencies that appear at precise differences, sums, and multiples of independent pulsation frequencies (Table~\ref{tab:freqs}). These combination frequencies are not comparable to calculations from stellar models and are important to identify and exclude from asteroseismic analyses. The amplitudes detected for the combination signals are expected to be much smaller than the independent pulsation mode amplitudes, and they are typically only detected for combinations of the highest-amplitude modes.  We find an exception in the presence of a significant peak at the sum of the frequencies of two low-amplitude modes, $f_4$ and $f_5$, suggesting that $f_4 + f_5$ may actually be an independent pulsation frequency that could improve our asteroseismic constraints. 
Still, there is less than a 0.5\% chance of an independent mode coinciding this precisely with the sum of any nine other pulsation frequencies.
We consider the risk of including a combination frequency in our model comparison much greater that the reward of an ostensibly better but possibly inaccurate fit. The peak at $f_4 + f_5$ would not have been adopted for exceeding our independent significance threshold regardless, and its inclusion as a combination frequency in our solution has a negligible effect on the period measurements for the independent modes. We do not consider the possibility that one of $f_4$ or $f_5$ may be a difference frequency involving the peak at $f_4+f_5$, as these individually have higher observed amplitudes and match the pattern of mode frequencies expected from nonlinear pulsation theory.

We note an interesting possibility to test the hypothesis that individual peaks are consistent with combination frequencies based on their relative amplitudes. \citet{Wu2001} provides analytical expressions for the amplitudes of combination frequencies that are based on a physical model for the nonlinear response of the stellar convection zone to the pulsations. This provides a framework for interpreting the amplitudes of combination frequencies in relation to their parent mode amplitudes that could constrain their spherical degrees, $\ell$, and azimuthal orders, $m$ \citep[e.g.,][]{Montgomery2005,Provencal12}.   Besides resolving the common $\ell$ ambiguity when comparing measured to model mode periods, identifying non-axisymmetric modes ($m\neq0$) would enable us to apply corrections to our measurements to recover $m=0$ period estimates. This would alleviate the systematic errors from assuming $m=0$ in the model fits, bringing the accuracy of our asteroseimic inferences closer to the level of precision that we currently achieve with the \emph{TESS} data. Tools to constrain mode identifications in this way for space-based photometry of pulsating white dwarfs are currently in development.

This collaborative first-light analysis from TASC WG8.2 has demonstrated the quality of the \emph{TESS} observations for measuring pulsations of DBV stars and the current state-of-the art of their interpretation. 
\emph{TESS} is continuing to observe new and known pulsating white dwarfs over nearly the entire sky, providing precise and reliable pulsation measurements for extensive asteroseimic study. Additional first-light papers from TASC WG8.2 on the pulsating PG\,1159 star NGC\,246 (Sowicka et al.) and an ensemble of DAV stars (Bogn{\'a}r et al.) are in preparation.

\begin{acknowledgements}
    We thank TASC WG8.2 for supporting this project and providing valuable feedback, especially D.~Kilkenny and R.~Raddi.
    We thank the anonymous referee whose comments helped to improve this manuscript.
    KJB is supported by an NSF Astronomy and Astrophysics Postdoctoral Fellowship under award AST-1903828.
    MHM acknowledges support from NSF grant AST-1707419 and the Wootton Center for Astrophysical Plasma Properties under the United States Department of Energy collaborative agreement DE-FOA-0001634.
    ASB gratefully acknowledges financial support from the Polish National Science Center under project No.\,UMO-2017/26/E/ST9/00703.
    ZsB acknowledges the financial support of the K-115709 and PD-123910 grants of the Hungarian National Research, Development and Innovation Office (NKFIH), and the Lend\"ulet Program of the Hungarian Academy of Sciences, project No.\,LP2018-7/2018.
    Support for this work was provided by NASA through the TESS Guest Investigator program through grant 80NSSC19K0378.
      This paper includes data collected with the \emph{TESS} mission, obtained from the MAST data archive at the Space Telescope Science Institute (STScI). Funding for the \emph{TESS} mission is provided by the NASA Explorer Program. STScI is operated by the Association of Universities for Research in Astronomy, Inc., under NASA contract NAS 5–26555.
    This work has made use of data from the European Space Agency (ESA) mission
    {\it Gaia} (\url{https://www.cosmos.esa.int/gaia}), processed by the {\it Gaia}
    Data Processing and Analysis Consortium (DPAC,
    \url{https://www.cosmos.esa.int/web/gaia/dpac/consortium}). Funding for the DPAC
    has been provided by national institutions, in particular the institutions
    participating in the {\it Gaia} Multilateral Agreement.
      
\end{acknowledgements}


\end{document}